\newif\ifpreprint
\preprintfalse
\ifpreprint
  \PassOptionsToClass{aip,apl,preprint}{revtex4-2}
  \newcommand{\figscale}{0.6}
\else
  \PassOptionsToClass{prx,reprint}{revtex4-2}
  \newcommand{\figscale}{1.0}
\fi
\documentclass[
  superscriptaddress,
  nofootinbib,
  floatfix,
  amsmath,
  amssymb,
]{revtex4-2}

\usepackage[T1]{fontenc}
\usepackage{braket}    
\usepackage{graphicx}
\usepackage{tikz}
\usepackage{xcolor}
\usepackage{siunitx}
\usepackage{hyperref}
\usepackage{makecell}
\hypersetup{
  colorlinks=true,
  linkcolor=black,
  citecolor=black,
  urlcolor=black,
  pdfborder={0 0 0}
}
\usepackage[normalem]{ulem}

\makeatletter
\def\@bibdataout@init{}%
\let\pre@bibdata\@empty
\makeatother

\newcommand{\baeight}{\(^{138}\text{Ba}^+\)}
\newcommand{\sreight}{\(^{88}\text{Sr}^+\)}
\newcommand{\cafour}{\(^{40}\text{Ca}^+\)}

\newcommand{\tpi}{\(\pi\)}

\newcommand{\tsig}{\(\sigma\)}

\newcommand{\tpsig}{\(\sigma^+\)}
\newcommand{\oneg}{\ensuremath{\ket{1_g}}}

\newcommand{\zerog}{\ensuremath{\ket{0_g}}}

\newcommand{\ketbra}[2]{\ket{#1}\!\bra{#2}}
\newcommand{\proj}[1]{\ketbra{#1}{#1}}

\begin{document}

\title{Remote entanglement need not be the bottleneck for~modular~trapped-ion~quantum~computing}

\author{Felix W. Knollmann}
\email{fwk@mit.edu}
\affiliation{Massachusetts Institute of Technology, Cambridge, Massachusetts 02139, USA}
\author{David P. Nadlinger}
\affiliation{Department of Physics, University of Oxford, Clarendon Laboratory, Parks Road, Oxford OX1 3PU, UK}
\author{John Blue}
\affiliation{Massachusetts Institute of Technology, Cambridge, Massachusetts 02139, USA}
\author{Sabrina M. Corsetti}
\affiliation{Massachusetts Institute of Technology, Cambridge, Massachusetts 02139, USA}
\author{Sam J. Bishop}
\affiliation{Massachusetts Institute of Technology, Cambridge, Massachusetts 02139, USA}
\author{Adam~R.~Martinez}
\affiliation{Department of Physics, University of Oxford, Clarendon Laboratory, Parks Road, Oxford OX1 3PU, UK}
\author{Jelena Notaros}
\affiliation{Massachusetts Institute of Technology, Cambridge, Massachusetts 02139, USA}
\author{Colin D. Bruzewicz}
\affiliation{Lincoln Laboratory, Massachusetts Institute of Technology, Lexington, MA, USA}
\author{Robert McConnell}
\affiliation{Lincoln Laboratory, Massachusetts Institute of Technology, Lexington, MA, USA}
\author{Isaac L. Chuang}
\affiliation{Massachusetts Institute of Technology, Cambridge, Massachusetts 02139, USA}

\date{\today}

\begin{abstract}
Modularity underpins classical computing; as quantum processors encounter limits on fabrication yield, reliability, and size, they will also need it acutely.
The bottleneck to linking modules is producing shared entanglement at sufficient rate, density, and fidelity.
Trapped ions hold the best demonstrated photonic links, yet they rely on bulky collection optics that cap how densely links can be packed, and remote entanglement operations trail local gates by two orders of magnitude in rate and fidelity.
We synthesize several enabling results – single-photon heralding, coherent recoil correction, projective distillation, and trap-integrated photonics – into one comprehensive architecture that substantially narrows this gap.
Single-photon heralding leads to linear scaling of success probability with detection efficiency, allowing compact integrated photonics to saturate the entanglement rate at a local-operation limit in dense, easy-to-parallelize channels.
Addressing its inherent error mechanisms at their source, we project a Bell-pair fidelity of 99.9\% at rates and densities compatible with fault-tolerant operations.
Remote entanglement then need not remain the bottleneck for modular trapped-ion computing; the limit shifts to the local operations that must improve regardless.
\end{abstract}
\maketitle

\section{Introduction}
\label{sec:intro}
A state-of-the-art surface-electrode ion trap quantum processor holds a few hundred ions with individual control~\cite{Ransford2025_Helios}, whereas architectures for factoring a 2048-bit RSA number call for \(10^4\) to \(10^7\) physical qubits~\cite{Gidney2025_FactorRSAMillionQubits,Webster2026_PinnacleArchitecture,Cain2026_ShorReconfigurableAtoms,Tripier2026_WalkingCatArchitecture,Zhou2025_TransversalArchitecture}. Closing this gap is challenging within a single trap: yield, heat dissipation, and wiring cap how large one processor can grow~\cite{Stick2006_IonTrapSemiconductorChip,Brown2021_MaterialsChallengesTrappedIon,Malinowski2023_Wiring1000QubitIonQC,Lekitsch2017_MicrowaveBlueprint,Bermudez2017_AssessingProgressTrappedIon}, while the slow physical operations of ion traps, such as shuttling, gates, and cooling, push algorithms toward parallelization that demands still more qubits~\cite{Cain2026_ShorReconfigurableAtoms,Tripier2026_WalkingCatArchitecture,Gidney2025_FactorRSAMillionQubits}. A modular solution, which links many processors rather than building one large one, allows link quality rather than the module size to determine whether the machine scales.

Useful quantum computation places certain demands on this link.
Following recent fault-tolerant architecture studies as a working baseline, whether based on transversal gates on surface codes~\cite{Zhou2025_TransversalArchitecture} or lattice surgery on quantum low-density parity-check (qLDPC) codes~\cite{Cain2026_ShorReconfigurableAtoms,Webster2026_PinnacleArchitecture,Tham2025_DistributedFaultTolerantMemories}, each module must supply tens to hundreds of Bell pairs at \(\ge\)99.9\% fidelity for every syndrome measurement cycle.
We assume that each module contains at least one surface-code patch or qLDPC code block.
Depending on whether that cycle is the aggressive \SI{1}{\milli\second} those architectures assume or the near-term \SI{55}{\milli\second} of a demonstrated operations layer~\cite{Ransford2025_Helios}, the target is \(10^4\)--\(10^6\) s\(^{-1}\) remote Bell pairs per module for the former and \(10^3\)--\(10^4\) s\(^{-1}\) for the latter (Table~\ref{tab:requirements_area}), delivered densely enough to efficiently interface with the approximately \si{\centi\meter\squared}-scale footprint of a single trap chip.
Simultaneously meeting all three demands, on rate, density, and fidelity, is the open problem.

No demonstrated approach meets these link demands simultaneously because the three are frustrated: improving one tends to degrade the others.
Trapped ions hold the records for heralded remote entanglement~\cite{Moehring2007_EntanglementAtomsDistance,Hucul2015_ModularEntanglement}, yet the best links reach only 97\% fidelity~\cite{Saha2025_TimeBinEntanglement,Main2025_DistributedQCNetworkLink} at rates up to \SI{250}{\per\second}~\cite{OReilly2024_ContinuouslyCooledEntanglement}, both roughly two orders of magnitude short of requirements, while local entangling operations already exceed 99.99\% at rates of \(10^4\) s\(^{-1}\)~\cite{Ransford2025_Helios,Hughes2025_SmoothGateIonQ}.
Standard two-photon heralding scales quadratically with photon collection efficiency~\cite{Barrett2005_DoubleHeraldingMatter,Simon2003_RobustLongDistance}, so proposals chase high collection efficiency with high-numerical-aperture (high-NA) lenses or optical cavities~\cite{Krutyanskiy2023_TrappedIon230m,Novakoski2025_PhotonicLinkedNetworks,Sutcliffe2025_HyperbolicFloquetDQEC,Carter2024_DualSpeciesImagingIonTrap,Kikura2026_PassiveQuantumInterconnects}; but larger collection optics are in direct tension with density requirements.
Entanglement distillation raises fidelity only by consuming additional raw pairs, taxing both rate and density~\cite{Bennett1996_PurificationNoisyEntanglement,Reichle2006_ExperimentalPurification,Nigmatullin2016_MinimallyComplexIonTraps}.

In this Perspective, we combine separately matured results in ways that create compounding benefits to relax this frustration rather than merely rebalancing it.
The key is to herald entanglement on a single photon~\cite{Cabrillo1999_CreationEntangledStates}, which makes the rate scale linearly rather than quadratically with detection efficiency.
This scaling allows even low-efficiency optics to reach the point where local operations, not photon collection, limit the distilled entanglement rate.
Compact integrated optics can then replace bulky lenses and cavities~\cite{Knollmann2024_IntegratedPhotonicStructures}, so that the size of the trap zone rather than the collection optic sets how densely channels can be tiled.
And fidelity can be decoupled from rate and density by avoiding deep distillation and instead addressing each error at its source with coherent recoil correction, leakage detection, and a single round of projective distillation~\cite{Apolin2025_RecoilInducedErrors,Campbell2008_ExtremePhotonLoss}.

We argue that this synthesis charts a practical path toward modular, fault-tolerant trapped-ion quantum computing, and analyze a quantitative example. We introduce the single-photon entanglement scheme (Sec.~\ref{sec:scheme}), show that the distilled rate saturates at its local-operation limit (Sec.~\ref{sec:rates}), identify the correction for each infidelity mechanism (Sec.~\ref{sec:fidelity}), and compare passive and optimized active photonic routing networks (Sec.~\ref{sec:architecture}, App.~\ref{app:networks}). For a full system assembled this way, we project a remote Bell-pair rate density of \(10^5\) s\(^{-1}\)cm\(^{-2}\) at 99.9\% fidelity with parallel all-to-all module connectivity (Sec.~\ref{sec:architecture}), using technologically realistic parameters.

\begin{table}[t]
    \centering
    \begin{ruledtabular}
    \begin{tabular}{l cc @{\hskip 1.5em} cc}
         & \multicolumn{2}{c}{Transversal \(O(d^2)\)}
         & \multicolumn{2}{c}{Surgery \(O(d)\)} \\[1pt]
        \hline
        \noalign{\vskip 2pt} 
        \(\tau_\mathrm{syndrome}\) (ms) & 1 & 55 & 1 & 55 \\
        Bell rate (s\(^{-1}\))
            & \(7.29\!\times\!10^{5}\) & \(1.33\!\times\!10^{4}\) & \(8.0\!\times\!10^{4}\)
            & \(1.5\!\times\!10^{3}\) \\[1pt]
        \hline
        \noalign{\vskip 2pt} 
        \multicolumn{5}{l}{\textit{This work (projected; Sec.~\ref{sec:projected_performance})}} \\[3pt]
        \quad Channels        & 850\textsuperscript{*}   & 16   & 94\textsuperscript{*}   & 2 \\
        \quad Area (cm\(^2\))  & 7.1\textsuperscript{*}  & 0.14 & 0.79\textsuperscript{*}  & 0.0168 \\[1pt]
        \hline
        \noalign{\vskip 2pt} 
        \multicolumn{5}{l}{\textit{State of the art}~\cite{OReilly2024_ContinuouslyCooledEntanglement}} \\[3pt]
        \quad Channels        & 2916  & 54   & 320  & 6 \\
        \quad Area (cm\(^2\))  & \(1.4\!\times\!10^{4}\) &  \(2.7\!\times\!10^{2}\) & \(1.57\!\times\!10^{3}\) & 29 \\
    \end{tabular}
    \end{ruledtabular}
    \caption{Per-module requirements for one intermodule two-qubit gate per code cycle on a large-scale error-corrected trapped-ion quantum processor (see App.~\ref{app:requirements} for details). \textsuperscript{*}Conservative: the faster operations required by a 1 ms syndrome measurement cycle would lower these values. The transversal scheme uses distance-\(27\) surface code patches that require one \(\tau_\mathrm{syndrome}\) measurement round with \(d^2 = 729\) Bell pairs~\cite{Zhou2025_TransversalArchitecture}. The surgery scheme uses a \(d\le20\) lifted product code~\cite{Cain2026_ShorReconfigurableAtoms} and code surgery~\cite{Swaroop2025_UniversalQLDPCAdapters} that requires \(d\) rounds of syndrome measurement with up to \(4d = 80\) Bell pairs each. This work assumes the distilled per-channel rate (\SI{858}{\per\second} at 99.9\% fidelity) and six-zone area (\SI{8.4e-3}{\centi\meter\squared}) of Sec.~\ref{sec:projected_performance} and projects \(\approx\!2000\times\) the entanglement rate density of the state-of-the-art baseline using the fastest demonstrated rate (\SI{250}{\per\second} at 94\% fidelity) and a collection-lens footprint (2.5~cm diameter)~\cite{OReilly2024_ContinuouslyCooledEntanglement,Carter2024_DualSpeciesImagingIonTrap}.}
    \label{tab:requirements_area}
\end{table}

\begin{figure}[!tb]
    \centering
    \includegraphics[width=\figscale\linewidth]{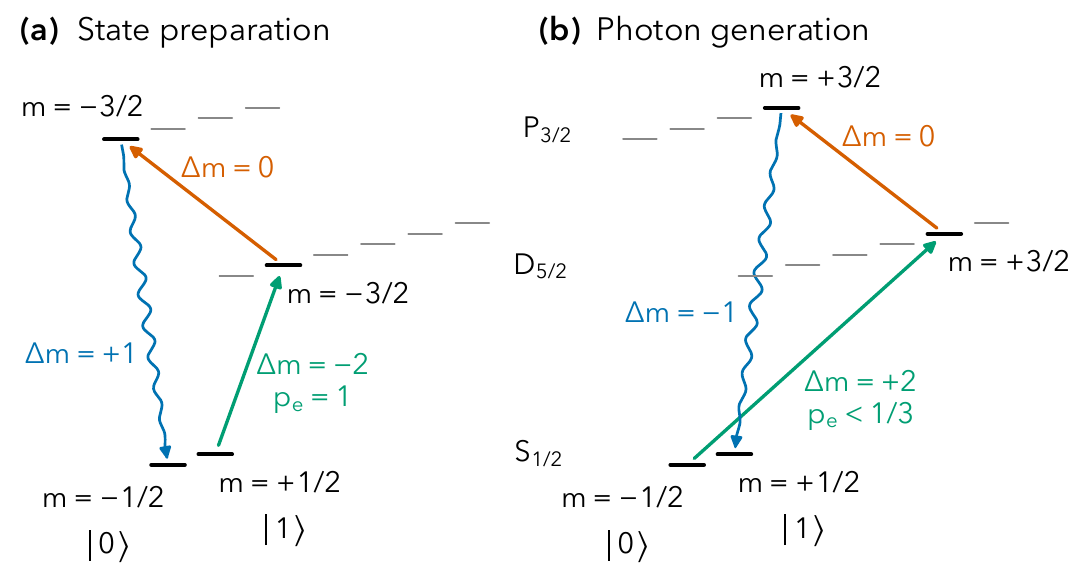}
    \caption{Energy level diagrams of commonly used \cafour, \sreight, and \baeight~ions that show the two-step excitation used for \textbf{(a)} state preparation and \textbf{(b)} entanglement without requiring circularly polarized light. An electric-quadrupole transition allows a change in angular momentum of \(|\Delta m|\)=2, such that a \tpi-polarized excitation to the stretched state of an excited level with a dipole-allowed decay back to the ground state leaves the ion with a net change of \(|\Delta m|\)=1 in the projection of angular momentum.}
    \label{fig:basic_scheme}
\end{figure}

\section{Entanglement scheme}
\label{sec:scheme} 
Remote entanglement is generated between one physical qubit in each of two modules. 
A weak excitation gives each qubit a small probability of emitting a photon; the photons are interfered at a central beamsplitter, so that detection of a single photon heralds entanglement between the two ions~\cite{Cabrillo1999_CreationEntangledStates}.
The modules are surface-electrode ion traps~\cite{Kielpinski2002_QCCDArchitecture,Chiaverini2005_SurfaceElectrodeArchitecture} with integrated optics for light emission and collection~\cite{Mehta2017_FocusingGratings,Knollmann2024_IntegratedPhotonicStructures}. 
The organization of \(M\) modules into a scalable network~\cite{Monroe2014_MUSIQC,DuanMonroe2010_QuantumNetworksTrappedIons,MonroeKim2013_ScalingIonTrapProcessor} is developed in Sec.~\ref{sec:arch_design} (Fig.~\ref{fig:architecture}). 

The physical process behind each herald is a single-photon, two-step scheme applicable to level structures similar to those of alkaline-earth ions with an \(\mathrm{S}\rightarrow{} \mathrm{D}\) quadrupole transition (Fig.~\ref{fig:basic_scheme}): each ion is prepared in one qubit state, driven on a narrow-line quadrupole transition to a metastable state with a finite probability $p_e$, and then excited to a short-lived state, from which it decays to the opposite qubit state by emitting a single photon. 
The intermediate state cleanly separates the photon-emitted and no-photon branches into distinct qubit states by angular-momentum selection rules.
The two-step structure spectrally separates the drive and emission light, so drive-light scatter can easily be suppressed by high-extinction filters at the detectors, and only linearly polarized light is required~\cite{Liu2026_LongLivedRemoteIonRepeaters}.
The same cycle is used for state preparation (Fig.~\ref{fig:basic_scheme}(a)), where a different narrow-linewidth laser frequency selectively empties the other qubit state without polarization purity requirements.
These qualities suit compact integrated photonics (Sec.~\ref{sec:architecture}), which can also reduce mode-matching infidelity~\cite{Knollmann2024_IntegratedPhotonicStructures}.

We prefer the single-photon entanglement scheme due to its linear detection efficiency scaling compared to the quadratic scaling of protocols using two-photon coincidence.
While the latter in principle implement ideal entanglement swapping, the single-photon scheme comes with intrinsic imperfections.
The raw entangled density matrix following a herald is
\begin{equation}
    \begin{gathered}
        \rho = (1-p_e)\proj{\Psi_{\chi,\phi}} + p_{e} \proj{11} \\
        \ket{\Psi_{\chi,\phi}} = \Big(\sqrt{\frac{1+\chi}{2}}K_A\ket{10}+e^{i\phi}\sqrt{\frac{1-\chi}{2}}K_B\ket{01}\Big)\ket{\zeta_A\zeta_B},
    \end{gathered}
\end{equation}
where the Bell state phase $\phi$ is set by the Mach–Zehnder interferometer of excitation and emission pathways~\cite{Cabrillo1999_CreationEntangledStates}, while the spin-dependent kick operators \(K_i\) capture entanglement between the qubit and motional state \(\ket{\zeta_i}\) of ion \(i\in\{A,B\}\) due to photon recoil~\cite{Apolin2025_RecoilInducedErrors}. A larger excitation probability $p_e$, desirable to increase the rate, leads to a larger double-excitation $\ket{11}$ component, while any imbalance of excitation and single-photon transmission probabilities between the two ions leads to an asymmetry between the $\ket{01}$ and $\ket{10}$ components as parametrized by \(\chi = (p_\mathrm{A}-p_\mathrm{B})/(p_\mathrm{A}+p_\mathrm{B}) \in [-1, 1]\).

To ensure that the delivered states are maximally entangled, the raw entanglement generation stage is followed by an error correction and projective distillation stage.
While spin-motion-entanglement errors can be severe in the single-excitation scheme~\cite{Slodicka2013_AtomAtomEntanglement}, a spin-dependent force can be applied after a herald success to disentangle spin and motion~\cite{Apolin2025_RecoilInducedErrors}.
Two raw Bell pairs are then combined in a node-local parity measurement to implement a single entanglement-distillation round, which removes the remaining double-excitation error and avoids sensitivity to interferometer phase and asymmetry drifts~\cite{Campbell2008_ExtremePhotonLoss,Kalb2017_EntanglementDistillationNV}.
A separate leakage-detection step flags ion decays to non-qubit states; if either distillation or leakage checks fail, the current candidate Bell pair is rejected.
Appendix~\ref{app:implementation} discusses this scheme in detail using~\sreight.
Because each error mechanism is addressed at its source, recoil correction, leakage detection and a single distillation round suffice to exceed 99.9\% distilled fidelity (Sec.~\ref{sec:fidelity}), avoiding the rate cost of deeper distillation (Sec.~\ref{sec:projected_performance}).

\begin{figure}
    \centering
    \includegraphics[width=\figscale\linewidth]{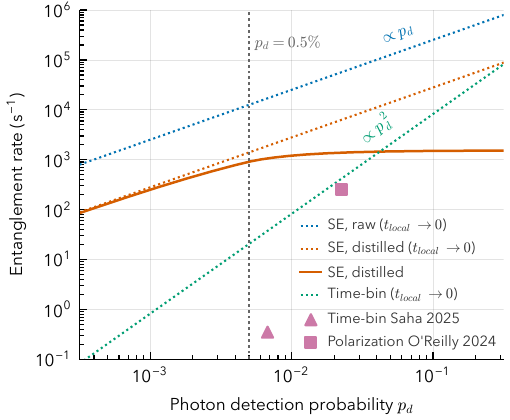}
    \caption{Heralded entanglement rate as a function of the photon detection probability \(p_d\) for \sreight, assuming the cycle times and local-operation timings in Table~\ref{tab:params}. \textbf{Dotted blue:} single-excitation raw heralded rate and \textbf{dotted orange:}\ distilled rate assuming instantaneous local operations (both at excitation probability \(p_e = 1/3\)). \textbf{Solid orange:} single-excitation rate under the parallelized model of Eq.~\ref{eq:t_distill_parallel}, with the rate set by the slowest of (Bell-pair, detection, CNOT) blocks and \(p_e\) optimized to maximize rate at each \(p_d\) using the parameters of Table~\ref{tab:params} but without accounting for buffer limitations. \textbf{Dotted green:} time-bin (two-photon coincidence) rate at \(p_e = 1\). The marked points denote the highest-rate demonstrations for polarization-encoded~\cite{OReilly2024_ContinuouslyCooledEntanglement} and time-bin-encoded~\cite{Saha2025_TimeBinEntanglement} entanglement of ions. The dashed vertical line marks the loss-tolerant working point \(p_d = 0.5\%\) used in the text.}
    \label{fig:rate_vs_detection}
\end{figure}

\section{Entanglement rate}
\label{sec:rates}
To deliver a distilled, high-fidelity Bell pair on a given channel, a fast loop repeats entanglement attempts of duration \(t_\mathrm{cycle}\), each heralding a raw Bell pair with probability \(p_\mathrm{herald}\). A slower correction loop then consumes two raw pairs to stochastically distill one high-fidelity pair with probability \(p_\mathrm{distill}\). The resulting rate (Fig.~\ref{fig:rate_vs_detection}) rises linearly with detection efficiency while collection is the bottleneck and then asymptotically approaches a local-operation limit.

The duration of the entanglement attempt \(t_\mathrm{cycle}\) is fundamentally limited by the linewidth of the excited state and the distance between the modules, as these determine the minimum time for state preparation and spontaneous emission and the decision branching time \(t_\mathrm{decision}\). For a given narrow-line \tpi-pulse duration \(t_\pi\), excited state lifetime \(\tau\), and latency of excitation \(t_\mathrm{excite}\) and photon detection \(t_\mathrm{detect}\), the cycle time is
\begin{equation}
    t_\mathrm{cycle}=t_\pi + t_e + 12\,\tau + t_\mathrm{decision} + t_\mathrm{excite} + t_\mathrm{detect}.
    \label{eq:t_cycle}
\end{equation}
The narrow-line drive time includes a \tpi-pulse for state preparation and \(t_e=\tfrac{2}{\pi}t_\pi\arcsin\!\sqrt{p_e}\) for excitation, where \(p_e\) is the desired excitation probability. The \(12\,\tau\) includes both time to completely clear the metastable level during state preparation and time for a photon to be emitted after excitation.

The herald probability per cycle \(p_\mathrm{herald}\) depends on the excitation probability \(p_e\), the branching ratio to the desired state \(\gamma\), the end-to-end detection probability \(p_d\), and the probability \(p_\mathrm{window}\) that spontaneous emission falls within the detector acceptance window. Since either ion A or B of the pair could emit a photon, the herald probability assuming perfect state preparation is
\begin{equation}
    \begin{aligned}
        p_\mathrm{herald}&\approx p_\mathrm{A} + p_\mathrm{B} \\
        p_\mathrm{i}&=p_{e,i}\,\gamma\,p_{d,i}\,p_\mathrm{window} \quad i\in\{A,B\}.
    \end{aligned}
    \label{eq:p_herald_se}
\end{equation}

\begin{figure}
    \centering
    \includegraphics[width=\figscale\linewidth]{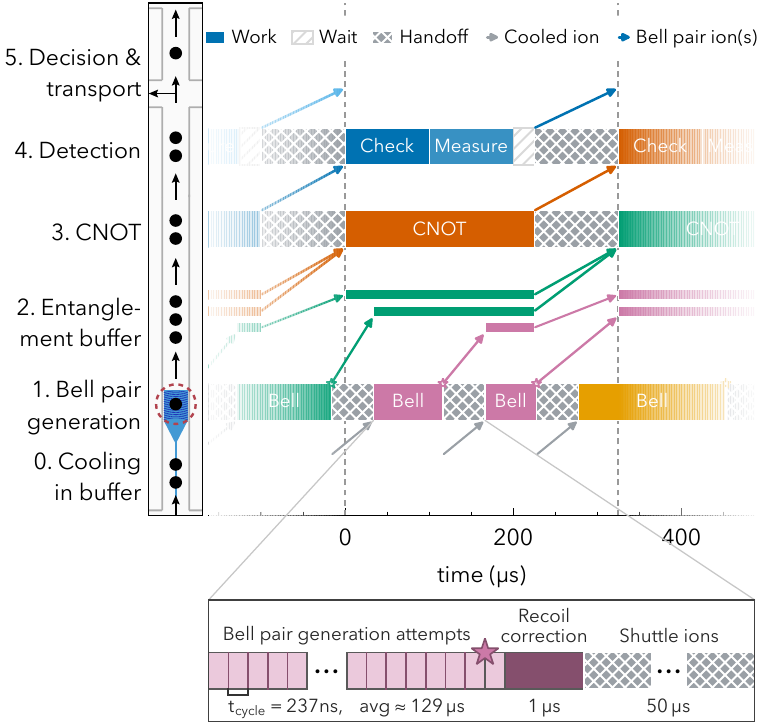}
    \caption{Pipeline for entanglement generation separated into blocks that can execute in parallel. Pre-cooled ions from a buffer move into the entanglement zone. After a successful herald, the recoil is corrected and consecutive ion pairs are buffered and then combined for a local CNOT. The ion pair undergoes leakage checks in the detection zone before the target ion is read out to determine the distilled state parity. A successfully distilled control ion exits the channel to be used while remaining ions are recycled to the starting buffer.}
    \label{fig:pipeline}
\end{figure}

One iteration of the correction loop, apart from the generation of two raw Bell pairs, also includes one CNOT gate, several single-qubit rotations, shuttling and merging of ion chains, and up to two detection windows at each module.
Single-qubit rotations are comparatively fast, so the duration is dominated by raw Bell-pair generation time \(t_\mathrm{Bell}\), detection time \(t_\mathrm{detect}\), the local CNOT duration \(t_\mathrm{CNOT}\), and per-handoff shuttling duration \(t_\mathrm{shuttle}\) (Sec.~\ref{sec:arch_design}).
Recoil correction takes comparable time to a single-qubit rotation due to the small phase space displacement required~\cite{Apolin2025_RecoilInducedErrors} and is neglected here.
We envision raw Bell-pair generation, the local CNOT, and the leakage-detected readout to be pipelined across dedicated zones (Fig.~\ref{fig:pipeline}) such that the steady-state distilled-pair rate is set by the slowest of these:
\begin{equation}
    t_\text{distill-cycle} = \max\left\{
        \begin{array}{l}
            2\,t_\mathrm{Bell}   + t_\mathrm{shuttle},   \\
            t_\mathrm{CNOT}      + t_\mathrm{shuttle},   \\
            2\,t_\mathrm{detect} + t_\mathrm{shuttle}
        \end{array}
    \right\}.
    \label{eq:t_distill_parallel}
\end{equation}
The first detection checks for one set of leakage events and the second combines further leakage checks with state readout of the CNOT target ion for distillation (App.~\ref{app:leakage}).
At low detection probability the Bell-pair block dominates and the distilled rate scales linearly with \(p_d\); once the local-operation blocks become comparable, the cycle time is pinned to the slowest local block and the rate depends on \(p_d\) only weakly, through the optimal \(p_e\) (Fig.~\ref{fig:rate_vs_detection}). 

Distillation succeeds with a probability \(p_\mathrm{distill}\) limited by the excitation probability \(p_e\) and the excitation and detection symmetry \(\chi\):
\begin{equation}
    p_\mathrm{distill} = \frac{1}{2}\,(1 - \chi^2)\,(1-p_{e})^2.
    \label{eq:dist_success}
\end{equation}
The average entanglement generation rate per parallel channel of \(R=p_\mathrm{distill}/t_\text{distill-cycle}\) can be guaranteed by buffering distilled Bell pairs before distribution.
The trade-off in \(p_e\) between \(p_\mathrm{distill}\) and \(p_\mathrm{herald}\) makes the distilled Bell-pair rate insensitive to \(p_{e}\) near the optimum. In the limit of vanishing local operation durations ($t_\mathrm{CNOT}$, $t_\mathrm{detect}$, $t_\mathrm{shuttle}$), the optimum \(p_{e}\) is 1/3, otherwise it lies lower. 

\subsection{Comparison with time-bin scheme}
\label{sec:timebin}
The natural point of comparison to the single-excitation scheme is a two-photon coincidence scheme in which each entanglement attempt drives both ions in two time bins (early and late). The photonic qubit is encoded in the bin of arrival, and the two photons are interfered at a linear-optics Bell-state analyzer~\cite{Barrett2005_DoubleHeraldingMatter,Saha2025_TimeBinEntanglement}. As the protocol no longer admixes a double-excitation error, the excitation probability can be unity. The extra rotations included in the time-bin cycle cost \(\tfrac{3}{2}\,t_\pi\) in addition to \(3\,t_\pi\) from state preparation and excitation, such that \(t_\mathrm{cycle}^\mathrm{TB} \approx \tfrac{9}{2}\,t_\pi + 18\,\tau + t_\mathrm{decision} + t_\mathrm{excite} + t_\mathrm{detect}\). The herald requires a detected photon from each ion within the acceptance window; a standard linear-optics analyzer resolves two of the four Bell states. The per-attempt herald probability is therefore
\begin{equation}
    p_\mathrm{herald}^\mathrm{TB} \approx \tfrac{1}{2}\,(p_e\,\gamma\,p_\mathrm{window}\,p_d)^2,
    \label{eq:p_herald_tb}
\end{equation}
quadratic in \(p_d\) and so penalized increasingly heavily as the collection efficiency drops (Fig.~\ref{fig:rate_vs_detection}). The photon-recoil correction introduced in~\cite{Apolin2025_RecoilInducedErrors} allows the delay between time bins to be chosen freely to optimize the entanglement rate, while prior demonstrations had to wait for an integer number of periods of the ion's motion between bins to suppress photon-recoil-induced errors~\cite{Saha2025_TimeBinEntanglement}. Other options for coincidence-based entanglement schemes include polarization- and frequency-based schemes~\cite{Barrett2005_DoubleHeraldingMatter}. However, these schemes are less suited to implementation with compact integrated optics. They suffer from polarization mixing challenges and difficulty correcting the recoil error in the frequency basis~\cite{Knollmann2024_IntegratedPhotonicStructures,Apolin2025_RecoilInducedErrors}. 

\begin{figure}[!tb]
    \centering
    \includegraphics[width=\figscale\linewidth]{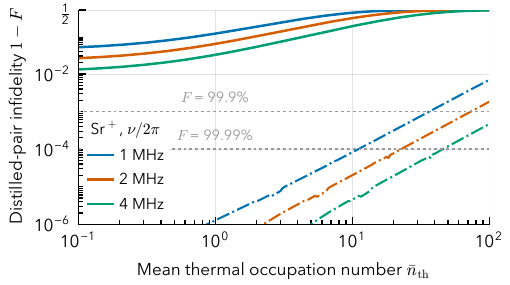}
    \caption{The Bell-state error \(1-F\) for \sreight~as a function of the mean motional occupation number \(\bar n_\mathrm{th}\), for three secular frequencies \(\nu/2\pi\in\{1,2,4\}\,\mathrm{MHz}\) and assuming a collection optic NA of 0.4; solid curves are the bare result, dot-dashed curves include the post-herald application of a spin-dependent displacement to correct photon recoil~\cite{Apolin2025_RecoilInducedErrors} and successful distillation (spin-dependent displacement and distillation assumed to be error-free).}
    \label{fig:fidelity_panels_motion}
\end{figure}

\section{Entanglement fidelity}
\label{sec:fidelity}
The correction and projective distillation loop described in Sec.~\ref{sec:scheme} is tailored to suppress errors intrinsic to the single-photon scheme; in this section, we discuss the errors remaining in the remote Bell pairs delivered to the QPU.
We treat only the errors particular to the entanglement scheme; general issues such as qubit dephasing and two-qubit gate or readout imperfections also affect the delivered Bell pairs, but must already be well-controlled in a working quantum computer. Further details are given in App.~\ref{app:outer_loop}.

\begin{table}[t]
    \centering
    \begin{ruledtabular}
    \begin{tabular}{l c c c}
        Error mechanism                 & Raw inf.                    & Residual inf.  & Proj.\,inf. \\
        \hline\\[-8pt]
        Double excitation               & \(p_e\)                           & 0 & 0\\[1pt]
        Phase drift                     & \(\sin^2(\frac{\phi_i}{2})\) & \(\sin^2(\frac{\phi_2-\phi_1}{2})\) & 0.007\%\\[1pt]
        Recoil effects                  & \multicolumn{2}{c}{see Fig.~\ref{fig:fidelity_panels_motion}} & 0.002\%\\[1pt]
        \makecell[l]{State preparation\\ (\(\epsilon_\mathrm{sp}\) left in \(\ket{1}\))} & \(\epsilon_\mathrm{sp} (1 - p_e)\)            & 0 & 0 \\[1pt]
        \makecell[l]{Decay to \(\ket{0}\)\\ (\(\epsilon_\mathrm{ret}\) of excited pop.)}  &  \(-p_e \epsilon_\mathrm{ret}/ 2\)  & \(\approx \frac{p_e}{1-p_e} \epsilon_\mathrm{ret}\) & 0.006\% \\[1pt]
        False herald                    & \(p_{\mathrm{false}} (1-p_e)^2\)            & \(\approx 6 p_e p_{\mathrm{false}}\) & 0.024\% \\
    \end{tabular}
    \end{ruledtabular}
    \caption{This table summarizes the Bell-pair infidelities before (raw) and after (residual) recoil correction and successful distillation, as derived in App.~\ref{app:distillation}. The projected infidelities are based on assumptions of Sec.~\ref{sec:projected_performance} and parameters from Table~\ref{tab:params}. Bell pair \(i\) has entangled-state phase \(\phi_i\), defined such that the target Bell state has \(\phi_i=0\). The raw decay to \(\ket{0}\) error actually reduces the bare \(p_e\) double-excitation error to \((1-\frac{\epsilon_\mathrm{ret}}{2})\,p_e\) (see App.~\ref{app:distillation}). Polarization impurity of the second excitation drive is the dominant example of returning to the initial state after excitation. False heralds due to detector dark counts, stray light, and network crosstalk have the same infidelity effect.}
    \label{tab:fidelity}
\end{table}

Table~\ref{tab:fidelity} summarizes the main error mechanisms affecting the raw Bell pairs and the distilled result. Double-excitation and state preparation errors are entirely suppressed by distillation (at the cost of a reduction in rate).
The distillation stage removes a phase common to both raw Bell pairs, but propagates the phase difference~\cite{Campbell2008_ExtremePhotonLoss}.
Long-term interferometric stability of the apparatus is thus not required, but across the two Bell pairs, a difference $\delta\phi$ in phase of the excitation pulses or along the collection pathways in either node decreases the fidelity as \(1-F=\sin^2(\delta\phi/2)\) (assuming \(\chi=0\)).
This insensitivity to long-term drift allows separate lasers to be used at each node as long as they are locked with sufficient short-term phase stability, and it dramatically reduces the need for active path-length stabilization. These points make the scheme attractive for complex, large-scale networks.

In general, the distillation circuit only rejects Z errors common to both Bell pairs; it doubles the sensitivity to individual Z errors.
One source of such errors is residual spin-motion entanglement due to photon recoil.
A post-herald correction using a simple spin-dependent displacement, as is standard in the trapped-ion toolbox, suffices to remove most of these errors for moderate ion temperatures (Fig.~\ref{fig:fidelity_panels_motion}).
For nonzero collection solid angles, the collection mode differs from a plane wave, so the first-order correction suggested here is only approximate~\cite{Apolin2025_RecoilInducedErrors}.

Following an excitation, a photon-emission path that returns the ion to its initial qubit state can produce a parity-correct but unentangled outcome, which the distillation circuit fails to reject.
This can occur due to polarization impurities.
Finally, as for any heralded entanglement generation scheme, false heralds due to photons not generated by the ions – whether from detector dark counts, uncontrolled scatter, or signal photon crosstalk in the optical routing fabric – can generate unentangled states, which are only partially rejected by distillation.
Appendix~\ref{app:distillation} treats these potential infidelity sources in more detail.

\begin{figure*}[ht!]
    \centering
    \includegraphics[width=1.0\linewidth]{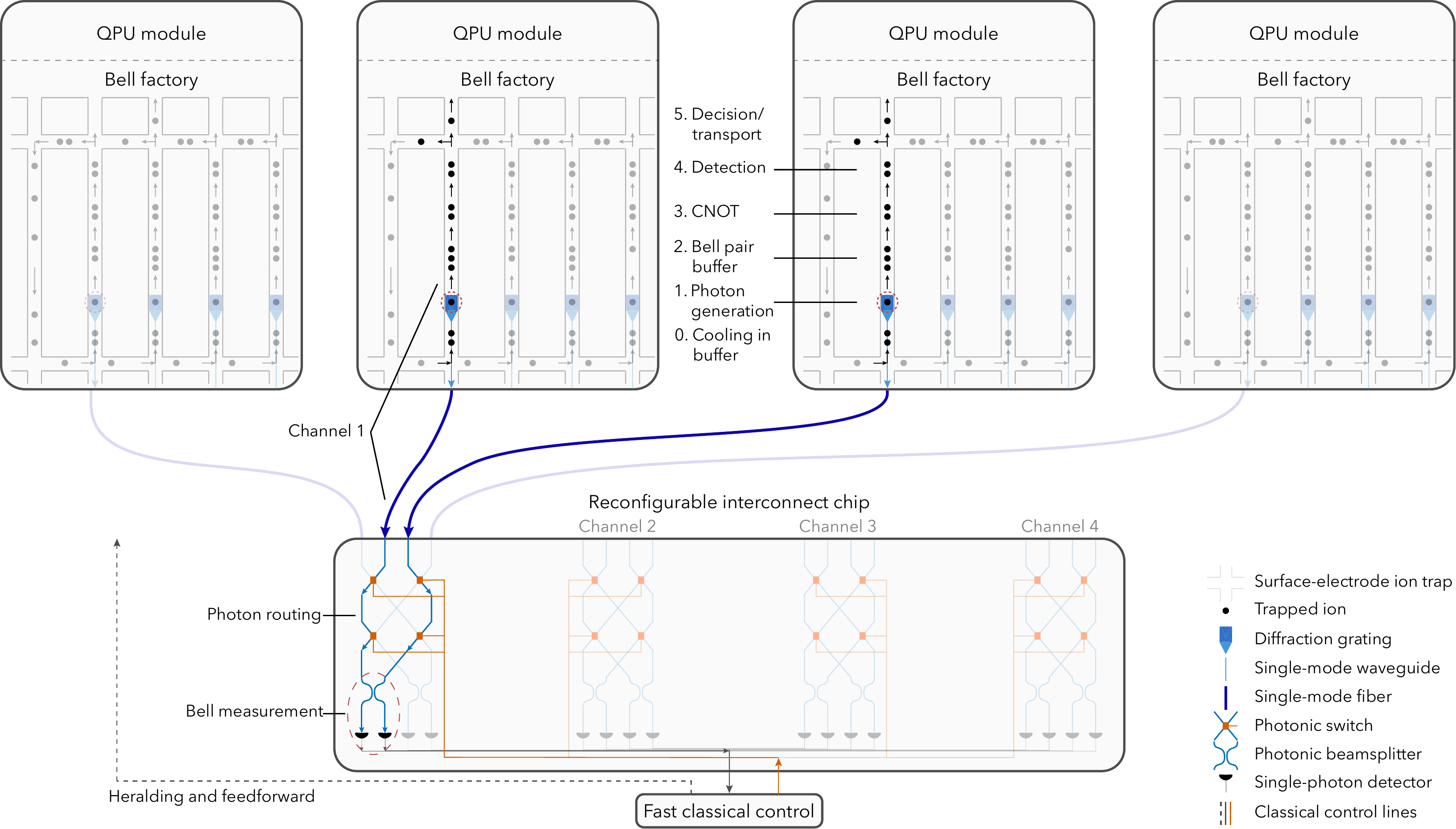}
    \caption{Remote entanglement generation architecture. Each quantum processing unit (QPU) is a module consisting of a microfabricated surface-electrode trap with integrated optics. The Bell factory produces and distills remote Bell pairs for intermodule operations. The photonic links consist of single-mode waveguides and fibers. A switch network enables all-to-all connectivity and passive beamsplitters provide the photon interference needed for entanglement. Detectors herald entangled states and a classical processing unit coordinates the entanglement process.}
    \label{fig:architecture}
\end{figure*}

\section{Architecture and system performance}
\label{sec:architecture}
This section assembles the entanglement scheme into a complete, scalable architecture and evaluates its end-to-end performance using demonstrated device parameters.

\subsection{Architecture}
\label{sec:arch_design}
Our Bell factory architecture scales to large numbers of modules while maintaining flexible connectivity, high entanglement rate, and high Bell-pair density. Fig.~\ref{fig:architecture} shows how local nodes are structured around the single-excitation entanglement scheme and associated fidelity corrections described in Sec.~\ref{sec:scheme}. Ions are moved along parallel channels following a QCCD architecture with dedicated zones for photon generation, gates, detection, cooling and entanglement buffers, and transport (Figs.~\ref{fig:architecture} and~\ref{fig:pipeline})~\cite{Kielpinski2002_QCCDArchitecture}. Raw Bell pairs are distilled within each entanglement channel to ensure that slow drifts affecting the entangled state phase are common mode. With transport zones shared over the entire Bell factory, each channel requires six QCCD zones and two junctions. Because the scheme's loss tolerance allows for compact, integrated collection and emission optics~\cite{Knollmann2024_IntegratedPhotonicStructures,Knollmann2026_FluorescenceCollectionGrating}, each zone is self-contained and channels can be densely tiled. The architecture parallelizes control across channels for transport, entanglement attempts, gates, and readout to minimize the individual control lines needed to operate many channels~\cite{Malinowski2023_Wiring1000QubitIonQC}. The central station is split by input channel into \(C\) independent photonic networks for photon interference and detection, which enable independent \(M\)-module connectivity maps on each of the \(C\) channels (Fig.~\ref{fig:architecture}). Multiple central stations could scale the system to arbitrarily large module numbers.

Both emission collection and drive-light delivery require similar channel-specific photonic networks to achieve the desired single-photon interference. The emission collection networks enable interference of arbitrary inputs and could be either passive symmetric multiports~\cite{Ainley2024_MultipartiteMultiNode} or active butterfly switch networks (App.~\ref{app:networks}); the choice depends on the component specifications and required connectivity density based on the trade-offs presented in Table~\ref{tab:photonic_networks}. On the drive side, each channel in a module requires one switch per wavelength to gate module participation in that entanglement cycle. In contrast to existing schemes, our scheme relaxes the extinction ratio requirements because the two-step excitation with the change in net angular momentum turns most imperfect-extinction errors into rate hits after leakage detection and distillation. This robustness to leaked state preparation and excitation light opens the door to compact and scalable integrated switches~\cite{Assumpcao2024_TFLNQuantumNodes,Dong2022_CMOSProgrammablePhotonics,Momenzadeh2025_ScalableIonAddressingPhotonicCircuits}. To minimize control requirements, a single modulator on each of the two excitation lasers can shape the pulses for state preparation and excitation. From that single source, the light can be fanned out using passive structures to each channel in each module with just the final participation-gating switch. App.~\ref{app:networks} explores these photonic networks in detail.

\subsection{Projected performance}
\label{sec:projected_performance}
We use the parameters defined in Table~\ref{tab:params} to evaluate the entanglement rate, fidelity, and density for a distilled \sreight--\sreight~Bell pair delivered to the edge of each QPU; mapping that pair onto QPU data qubits is not included in the budgets below. Our analysis uses \sreight~because it has the highest branching ratio \(\gamma\) back to the ground state of similar ions and achieves a given entanglement rate with the least optical power (App.~\ref{app:power}).

First, we estimate the entanglement rate. Using the Table~\ref{tab:params} parameters, the single-excitation cycle time is \SI{237}{\nano\second}, so the attempt rate is \SI{4.2e6}{\per\second}. The latencies \(t_\mathrm{excite},\,t_\mathrm{detect}\) assume a distance \(<8\) m between modules and the speed of light in optical fiber. We choose a 75 ns \(t_\pi\) to balance entanglement rate and optical power. It requires 4.4 mW for a square pulse using a 2 \textmu m Gaussian beam (see App.~\ref{app:power} for the trade-off). With routing and detection losses based on the current state of the art (\(-4\,\mathrm{dB}\)) and predicted collection efficiency from grating couplers covering a numerical aperture of 0.4 (\(-19\,\mathrm{dB}\)), the end-to-end detection efficiency could be 0.5\% (\(-23\,\mathrm{dB}\)) (Table~\ref{tab:feasibility}), conservative in comparison with the 2--3\% end-to-end detection efficiency demonstrated for lens-based collection~\cite{OReilly2024_ContinuouslyCooledEntanglement,Stephenson2020_HighRateOxfordSr}. 
For our parameters and including a finite buffer capacity \(N_\mathrm{buf}=4\), the optimum excitation probability is \(p_e=0.198\). Under these assumptions, \(p_\mathrm{herald} = 0.18\%\) and the average rate of heralds is \SI{7.7e3}{\per\second}, which is similar to local operation rates in trapped ions (Table~\ref{tab:params}).

Our optimal \(p_e=0.198\) yields \(p_\mathrm{distill}=0.32\). The three pipelined blocks of Eq.~\ref{eq:t_distill_parallel} have comparable durations of 359, 325, and 300 \textmu s. Since the Bell-pair generation is slowest, the average distilled rate is \(p_\mathrm{distill}/\SI{359}{\micro\second}=\SI{896}{\per\second}\) per channel if using an infinite buffer. Fig.~\ref{fig:rate_vs_detection} shows how once local operations become limiting, the rate becomes independent of further improvements in collection or detection efficiency, setting an upper bound for the target raw Bell-pair generation rate. Our parameters sit just below this limit, but decreasing \(t_\pi\) by increasing the narrow-line laser power pushes the rate into the local-operations plateau (Fig.~\ref{fig:rate_vs_power}). Without pipelining, the sum of the block times would set the distilled rate and reduce it to \SI{327}{\per\second}, quantifying the rate advantage of the pipelined architecture.

Next, we determine the Bell pair density. The delivered per-channel rate is set by the finite entanglement buffer: a four-ion buffer locks the distillation cadence to the 325 \textmu s CNOT block instead of the 359 \textmu s Bell-pair block, giving an unskipped rate of \(p_\mathrm{distill}/\SI{325}{\micro\second}=\SI{990}{\per\second}\); a Monte-Carlo-modeled 13.3\% buffer-starvation skip fraction then reduces this to a net \SI{858}{\per\second} per channel (assuming the buffer-to-buffer swaps around the entanglement zone can be done at half of the usual inter-zone handoff time \(t_\mathrm{shuttle}\), see Fig.~\ref{fig:pipeline}). Assuming the demonstrated square zones of side length 375 \textmu m from~\cite{Delaney2024_ScalableMultispeciesTransport} and an amortized 6 zones per channel (cooling buffer, entanglement, entanglement buffer, gate, readout, transport), then each channel holds up to 13 ions at steady state and occupies 0.84 mm\(^2\) for an entanglement rate density of \SI{1.0e5}{\per\second}cm\(^{-2}\) of Bell factory. Future devices will achieve higher densities since the full square grid of junctions used in~\cite{Delaney2024_ScalableMultispeciesTransport} is not required and zones will become more compact. The entanglement rate is decoupled from the density at our parameters because the 0.4 NA integrated collection optic covers only 0.0015 mm\(^2\) of the 0.14 mm\(^2\) per QCCD zone. The per channel rate (\SI{858}{\per\second}) and area (0.84 mm\(^2\)) give rise to the values in Table~\ref{tab:requirements_area}, which will improve as future devices increase shuttling rates and zone densities.

Finally, we examine the fidelity limitations. We choose a phase deviation between consecutive Bell pairs $\sigma_{\delta\phi}$ based on phase noise spectral densities reported for $\SI{10}{\metre}$-scale fiber patch cords in a laboratory environment (see App.~\ref{app:phase-stability}).
The relative path length could be further stabilized by light spectrally separated from the emitted photons, as has already been demonstrated over 10~km path length~\cite{Liu2026_BayesianPhaseStabilization} using well-established techniques for active cancellation of fiber phase noise~\cite{Minar2008_PhaseNoiseLongFiber,Grosche2009_OpticalFrequencyTransfer}.
Alternatively, discarding a raw pair whose partner is not heralded within a set time would bound the phase excursion at a small rate cost.
Doppler cooling in the buffer suffices to bound the motion after a successful herald to the assumed thermal state (App.~\ref{app:motion}). Of the dephasing errors where photon emission returns the ion to its initial qubit state, the least avoidable mechanism is polarization impurity on the second excitation step. Keeping the distilled fidelity above \(F_\mathrm{target}\) requires \(\varepsilon^2 \lesssim 2(1-F_\mathrm{target})(1-p_e)/(p_e\,\gamma)\), where \(\varepsilon^2\equiv I_{\sigma^-}/I_\pi\) (App.~\ref{app:distillation}). We choose an intensity ratio less than that already achieved at 674 nm using an integrated grating emitter~\cite{Mehta2017_FocusingGratings}. Other dephasing mechanisms are negligible in practice: population left in the metastable state before excitation is removed by deshelving after state preparation, and off-resonant excitation on the first step is suppressed by pulse shaping and beam geometry at a modest \(\sim\)20 G field, and double excitation is limited by the branching ratio, which eases the requirements on the second-excitation-step pulse length~\cite{Crocker2019_HighPuritySinglePhotons}. 

Recoil correction, distillation, and leakage checks remove the scheme's intrinsic errors, leaving only technical residual errors. At the assumed \(\sigma_{\delta\phi}\), \(\bar{n}\), and \(\varepsilon^2\) (Table~\ref{tab:params}), the link phase, motional, and polarization infidelities are 0.007\%, 0.002\%, and 0.006\% after distillation. The dominant term is a \(0.024\%\) false-herald infidelity from two detectors, assuming either an active network with high-extinction switches or a passive network of two modules (App.~\ref{app:networks}). Summed linearly, these errors bound the delivered fidelity to \(F\approx99.96\%\). Because each of these errors could be improved by technical choices – for instance, superconducting nanowire single-photon detectors (SNSPDs) offer much lower dark-count rates (\(R_\mathrm{dc}\ll0.01\) s\(^{-1}\))~\cite{Wollman2017_UVSNSPDs} – the achievable fidelity is \(F>99.9\%\) and set by the QPU's local operations rather than by the remote entanglement scheme.

\begin{table}[t]
    \centering
    \begin{ruledtabular}
    \begin{tabular}{l l l}
        Symbol & Description & Value \\
        \hline
        \multicolumn{3}{l}{\textit{Atomic / optical} (\sreight)} \\
        \(t_\pi\)              & Narrow-line \(\pi\) time                              & 75 ns \\
        \(\tau\)               & P\(_{3/2}\) lifetime~\cite{Pinnington1995_LifetimesSrBa} & 6.63 ns \\
        \(t_\mathrm{decision}\)& Decision latency                                      & 20 ns \\
        \(t_\mathrm{excite},\,t_\mathrm{detect}\) & One-way latency                & 20 ns \\
        \(\gamma\)             & Branching ratio~\cite{Zhang2016_BranchingRatios88Sr}  & 0.94 \\
        \(p_\mathrm{window}\)  & \(4\tau\) acceptance window                           & 0.98 \\
        \(p_d\)                & Photon detection probability                          & 0.005 \\
        \(\chi\)               & Efficiency asymmetry, $\chi \in [-1, 1]$              & 0 \\
        \(\sigma_{\delta\phi}\)        & Link phase uncertainty (see App.~\ref{app:phase-stability}) & $< 0.02$ rad\\
        \(R_\mathrm{dc}\)     & Dark-count rate\footnote{Hamamatsu C11202-050}        & 7 s\(^{-1}\)\\
        \(\bar{n}\)            & Motion thermal state                                  & 10 \\
        \(\nu\)                & Trap frequency                                        & \(2\pi\times2\) MHz \\
        \(\varepsilon^2\)      & Intensity ratio \(I_{\sigma^-}/I_\pi\)~\cite{Mehta2017_FocusingGratings} & 0.05\%\\
        \hline
        \multicolumn{3}{l}{\textit{Local operations}} \\
        \(t_\mathrm{detect}\)  & Fluorescence detection~\cite{Gaebler2021_MidcircuitMeasurement} & 100 \textmu s \\
        \(t_\mathrm{CNOT}\)    & Local CNOT~\cite{Hughes2025_SmoothGateIonQ}            & 225 \textmu s \\
        \(t_\mathrm{shuttle}\) & Per-handoff shuttle / merge~\cite{Delaney2024_ScalableMultispeciesTransport}  & 100 \textmu s \\
        \(N_\mathrm{buf}\)     & Buffer capacity                                       & 4 \\
        \(d_\mathrm{zone}\)    & QCCD zone side length~\cite{Delaney2024_ScalableMultispeciesTransport} & 375 \textmu m \\
        \hline
        \multicolumn{3}{l}{\textit{Scheme-dependent}} \\
        \(p_e\) (SE)           & SE excitation probability                             & 0.198 \\
        \(p_e\) (TB)           & TB excitation probability                             & 1 \\
        \hline
        \multicolumn{3}{l}{\textit{Derived cycle times}} \\
        \(t_\mathrm{cycle}\)   & SE cycle, Eq.~\ref{eq:t_cycle}                        & 237 ns \\
        \(t_\mathrm{cycle}^\mathrm{TB}\) & TB cycle                                    & 517 ns \\
    \end{tabular}
    \end{ruledtabular}
    \caption{Input parameters of the entanglement rate, fidelity, and density models used throughout.
    Values without citations are assumptions and derived parameters. Approximate operation times are from state-of-the-art demonstrations. Symbols match those defined in Sec.~\ref{sec:rates},~\ref{sec:fidelity},~\ref{sec:projected_performance}, App.~\ref{app:distillation}, and App.~\ref{app:phase-stability}. SE = single-excitation, TB = time-bin.}
    \label{tab:params}
\end{table}

\section{Discussion}
\label{sec:discussion}
Having projected the rate, density, and fidelity, we situate the architecture by comparing it to existing trapped-ion interconnects, analyzing its application to other schemes and platforms, determining its feasibility using already-demonstrated components, and examining its implications.

\subsection{Comparison to prior work}
\label{sec:comparison}
The records for remote entanglement of \SI{250}{\per\second} and \(\le\)97\% fidelity are held by small-scale demonstrations with high-NA lenses and up to a few ions per node with no easy path to scaling channel density~\cite{OReilly2024_ContinuouslyCooledEntanglement,Main2025_DistributedQCNetworkLink,Saha2025_TimeBinEntanglement}. Other proposals for increasing trapped-ion interconnect rates fall into two categories: increasing end-to-end collection efficiency using optical cavities~\cite{Barrett2005_DoubleHeraldingMatter} or using quantum memories to circumvent photon coincidence requirements~\cite{Duan2001_AtomicEnsembleComm}. Optical cavities could achieve similar QPU rates but require substantial volume and can introduce infidelity mechanisms~\cite{Novakoski2025_PhotonicLinkedNetworks,Sutcliffe2025_HyperbolicFloquetDQEC,Kikura2026_PassiveQuantumInterconnects}. Quantum memories such as silicon vacancy centers in nanophotonic cavities promise similar \(p_d^2\rightarrow p_d\) scaling improvement to the single-excitation scheme but face challenges of lower-fidelity operations in solid-state qubits and coherent single-photon frequency conversion~\cite{Bhaskar2020_MemoryEnhancedComm}. A related single-photon ion link by Liu et al.~\cite{Liu2026_LongLivedRemoteIonRepeaters} heralds entanglement in the same loss-tolerant regime, but does not separate the photon-emitted and no-photon branches into distinct qubit states. The loss tolerance of our architecture (Fig.~\ref{fig:rate_vs_detection}) removes the need for optical cavities or quantum memories, enables dense scaling of parallel channels, and eases the engineering requirements for the collection optic and photonic network.

\subsection{Applicability to other platforms}
\label{sec:applicability}
While integrated optics are well-matched to the loss tolerance, parallelism, and density afforded by this scheme, our architecture also works with bulk optics or metalenses~\cite{Lim2026_TrapIntegratedMetalens} and fiber or free-space photonic networks. Our ion of choice, \sreight, is already a common networking ion~\cite{Main2025_DistributedQCNetworkLink,Inlek2017_MultispeciesNode} and functions as a sympathetic coolant to the barium ions commonly used as data qubits. It would be even better mass-matched to \(^{89}\)Y\(^+\)~\cite{Gilbreth2026_YttriumIonQIP}. The handoff between the networking qubit and the data qubits lies outside our scope.

If a given platform has access to high end-to-end photon collection, then the time-bin implementation of our scheme becomes competitive in rate. Figure~\ref{fig:rate_vs_detection} shows that local operation rates become the determining factor of when the cross-over happens. Recoil-correction enables high attempt rates and the cavity complexity is minimized since it needs to collect photons of only a single frequency and polarization. Unlike in the single-excitation case, the fast qubit rotation fidelity and off-resonant excitation suppression directly contribute to the entangled state fidelity in the time-bin approach. Because of this sensitivity and finite local operation times, the achievable rate will fall below the dotted line of Fig.~\ref{fig:rate_vs_detection}.

In principle, the entanglement scheme, distillation, and recoil correction all transfer directly to neutral atoms; however, recoil heating makes implementation with neutral atoms less practical. Tweezer-trapped atoms typically have Lamb–Dicke parameters 10\(\times\) higher than those of trapped ions, and the \(\eta^2\) scaling implies 100\(\times\) higher recoil heating. Combined with the worse recoil correction at higher \(\eta\)~\cite{Apolin2025_RecoilInducedErrors}, the cooling requirement for high-fidelity two-qubit gates~\cite{Petrosyan2017_RydbergDarkStateGate,Pagano2022_RydbergCZErrorBudget}, and difficulty of cooling neutral atoms while maintaining internal-state qubit coherence, this scheme appears less suitable for neutral atoms. Developments in the control of circular Rydberg states or cycling atoms to amortize recooling time could shift this calculus~\cite{Holzl2024_CircularRydbergQubits,Li2024_NeutralAtomCavityInterconnects}. 

\subsection{Technical feasibility}
\label{sec:feasibility}
The architecture requires a working QCCD quantum computer at each QPU such that ion rearrangement, readout, and local qubit operations all perform well enough not to limit the fidelity of our protocol~\cite{Bruzewicz2019_TrappedIonProgressChallenges}. The Zeeman qubits we use within the Bell factory require a combination of magnetic field stability and dynamical decoupling sufficient that the coherence time does not limit the distilled Bell-pair fidelity. Beyond these standard operations that have been demonstrated in commercial platforms~\cite{Ransford2025_Helios,Hughes2025_SmoothGateIonQ}, the scheme requires sophisticated timing and decision branching at nanosecond timescales across a network of modules as shown in~\cite{Stephenson2020_HighRateOxfordSr}. For the highest rates, the optical power required becomes substantial; however, the trade-off between optical power and channel rate is favorable (App.~\ref{app:power}). At our working point, a 2\(\times\) slower channel requires 20\(\times\) less optical power.

Most of the key integrated-optics metrics that the architecture requires (low-loss routing, balanced splitting ratios, focusing optics, and switching) have already been demonstrated individually~\cite{Moody2022_RoadmapIntegratedQuantumPhotonics}. Bell factories on \(2^3\) modules could achieve 4 dB detection and routing loss using amorphous Al\(_2\)O\(_3\) single-mode waveguides and commercially available fibers and detectors (Table~\ref{tab:feasibility}). Beamsplitter ratios better than 49:51 are routine using multi-mode-interferometer (MMI) splitters at 405 nm~\cite{Mashayekh2023_VisiblePICLifeScience}. Designs for focused-beam-emitting grating couplers have achieved 2 dB mode-matching loss with effective NA\(>\)0.4 (\(<\)14 dB solid angle limit) in dual layer Si\(_3\)N\(_4\), which could be replicated in dual-layer Al\(_2\)O\(_3\)~\cite{Knollmann2024_IntegratedPhotonicStructures,Knollmann2026_FluorescenceCollectionGrating}. Finally, a Mach--Zehnder mesh operated at 5 K has demonstrated \(>\)10 MHz bandwidth and 30 dB extinction ratio for 737-nm light, which meets the switching requirements for the excitation lasers~\cite{Dong2022_CMOSProgrammablePhotonics}; similar devices have already demonstrated control of a Ca\(^+\) qubit at 729 nm~\cite{Hogle2023_ScalablePhotonicModulators}. The only component whose architecture requirements exceed demonstrated performances is the high-extinction ratio, fast switch at 408 nm; the closest demonstrated device achieves a \(>\)10 dB extinction ratio at 420 nm~\cite{Castillo2026_UVAluminaModulator}. Developing such switches could unlock simultaneous entanglement attempts on all arbitrary module pairs to increase the system entanglement rate by a factor of up to \(\frac{M}{2}\). To meet the architecture's phase stability requirement between consecutive Bell pairs, switches should be driven push--pull, which cancels the common-mode phase imparted on the routed light and leaves a residual proportional to the fractional mismatch between the two arms' responses. 

\begin{table}[t]
    \centering
    \begin{ruledtabular}
    \begin{tabular}{l l l}
        Device & \(\lambda\) (nm) & Loss (dB) \\
        \hline
        \multicolumn{3}{l}{\textit{Routing loss}}\\
        Waveguide transmission~\cite{Kwon2024_MultiSiteOpticalAddressing}   & 369 & 1 cm \(\times\) 1.35 cm\(^{-1}\) \\
        Fiber transmission\footnote{Thorlabs PM-S405-XP}                    & 405 & 4 m \(\times\) 0.05 m\(^{-1}\) \\
        Waveguide-fiber coupler~\cite{Lin2025_AluminaUVEdgeCoupler}         & 405 & 2 \(\times\) 0.29  \\
        MMI transmission~\cite{Mashayekh2023_VisiblePICLifeScience}         & 405 & 3 \(\times\) 0.09  \\
        SPAD quantum efficiency\footnote{Hamamatsu C11202-050}              & 408 & 1.6  \\
        \hline
        \multicolumn{3}{l}{\textit{Collection loss}}\\
        Solid angle (NA = 0.4)  & 408 & 14  \\
        Polarization overlap    & 408 & 3  \\
        Mode matching~\cite{Knollmann2026_FluorescenceCollectionGrating}           & 408 & 2  \\
        \hline
        End-to-end loss         & 408 & 23\\
    \end{tabular}
    \end{ruledtabular}
    \caption{Photonic parameters taken from demonstrations and designs in the literature for calculating expected end-to-end photon detection efficiency for \sreight. The 0.29 dB transverse-electric-mode (TE-mode) edge coupler used a custom fabrication process~\cite{Lin2025_AluminaUVEdgeCoupler}; in a standard process, 3.53 dB has been achieved~\cite{Du2024_AluminaFiberCoupler}. The overlap of \tsig-photon dipole emission into a linear polarization mode introduces a 3 dB collection loss~\cite{Knollmann2026_FluorescenceCollectionGrating}. SPAD = single-photon avalanche diode.}
    \label{tab:feasibility}
\end{table}

We expect faster ion transport~\cite{Munoz2026_TransportExcitationBenchmarking} and more compact zones to increase the entanglement rate and density of future large-scale fault-tolerant devices and thus decrease the required channel numbers and Bell factory area. If the high-fidelity two-qubit gates used in distillation require ground-state cooling, the ions can be sympathetically cooled between leakage checks and distillation~\cite{Larson1986_SympatheticCooling}. However, since composite pulse sequences are known to address motion errors in single-qubit gates and recent 99.99\% fidelity two-qubit gates operate at the Doppler limit, it seems possible to skip this step~\cite{Hughes2025_SmoothGateIonQ,Wimperis1994_CompositePulses,Cummins2003_CompositeRotations,Timoney2008_ErrorResistantGates}. The cost difference in rate, fidelity, and space between local and remote two-qubit gates invites architectural exploitation using cache storage and memory/processor hierarchies~\cite{Thaker2006_QuantumMemoryHierarchies,Webster2026_PinnacleArchitecture,Cain2026_ShorReconfigurableAtoms,Tripier2026_WalkingCatArchitecture,Gidney2025_FactorRSAMillionQubits}.

\subsection{Outlook}
\label{sec:outlook}
Near our working point, the pipelined structure and its additive time floors mean no parameter enters the rate linearly, creating a broad region of high rate. While, with our deliberately conservative parameters, Bell-pair generation rather than local operations limits the per-channel rate, even perfect collection (\(p_d=1\)) would raise it by at most 1.8\(\times\). This robustness, together with the trap-zone-limited density, local-operation-limited fidelity, and straightforward multiplexing, makes the architecture practical. These dense interconnects with high rates and fidelities open up the design space of modular trapped-ion quantum computing, and invite systematic study of the optimal module size, composition, and specialization. Beyond distributed quantum computing, the same scheme could serve a range of quantum-networking applications, including quantum-enhanced long-baseline interferometry~\cite{Stas2026_NonLocalInterferometry,Gottesman2012_LongerBaselineTelescopes,Khabiboulline2019_OpticalInterferometryQuantumNetworks}, entangled clock networks~\cite{Komar2014_QuantumNetworkClocks,Nichol2022_EntangledOpticalClocks}, blind quantum computing~\cite{Fitzsimons2017_BlindQuantumComputing}, device-independent quantum key distribution~\cite{Nadlinger2022_DeviceIndependentQKD}, and quantum-computing-enhanced sensing~\cite{Khan2025_QuantumComputationalSensing,Allen2025_QuantumComputingEnhancedSensing,Giovannetti2011_AdvancesQuantumMetrology,Degen2017_QuantumSensing}.

\begin{acknowledgments}
    We acknowledge Gavin West, John Chiaverini, Ethan Clements, and Adrian Menssen for helpful discussions, and thank Tenzan Araki for comments on the manuscript.  Support in part is acknowledged from the NSF Frontier Center for Ultracold Atoms (grant number PHY-2317134). J.B. acknowledges support from the MIT Center for Quantum Engineering--Laboratory for Physical Sciences Doc Bedard Fellowship. This material is based upon work supported by the U.S. Department of Energy, Office of Science, National Quantum Information Science Research Centers, Quantum Systems Accelerator (Award No. DE-SCL0000121). This material is based upon work supported by the Department of Energy under Air Force Contract No. FA8702-15-D-0001 or FA8702-25-D-B002. Any opinions, findings, conclusions or recommendations expressed in this material are those of the author(s) and do not necessarily reflect the views of the Department of Energy.
\end{acknowledgments}




\newpage

\appendix

\section{Trapped-ion implementation details}
\label{app:implementation}
\begin{figure}[ht]
    \centering
    \includegraphics[width=\figscale\linewidth]{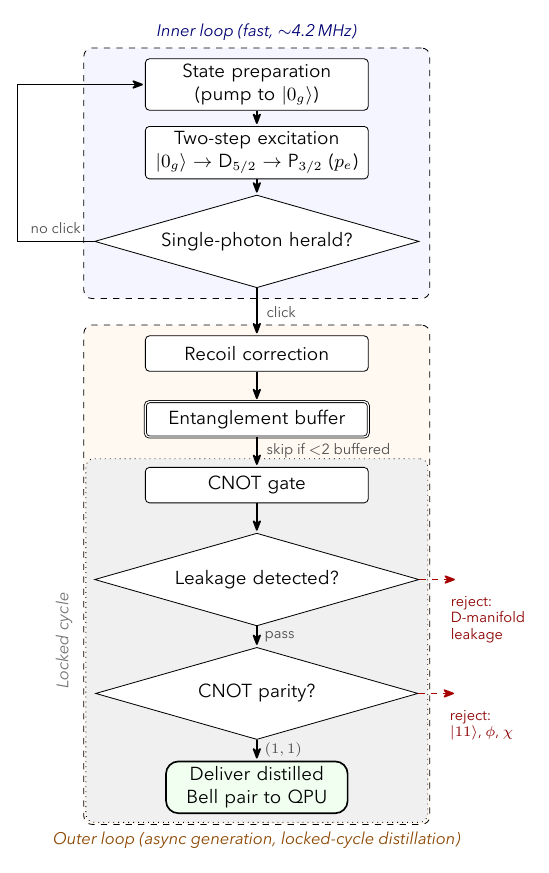}
    \caption{Control flow of the nested entanglement protocol (App.~\ref{app:implementation}).
    The fast \emph{inner loop} repeats state preparation and two-step excitation until a
    single-photon herald succeeds; the detector that clicked fixes the Bell-state sign.
    Each heralded Bell pair enters the \emph{outer loop}, which corrects photon-recoil spin--motion
    entanglement, pairs two raw Bell pairs in a buffer, and distills them in a locked cycle of two
    325 \textmu s steps using a CNOT and parity check with leakage detection (Fig.~\ref{fig:pipeline}). Dashed red paths mark
    the errors each check converts into rate loss rather than infidelity: distillation rejects
    double excitation \(\ket{11}\) and collection asymmetry $\chi$ and fixes the herald phase $\phi$;
    leakage detection rejects population that has escaped to the D manifolds.}
    \label{fig:loop_protocol}
\end{figure}

This appendix works through a concrete realization of the entanglement scheme in the relatively simple level structures of \cafour, \sreight, and \baeight, with the full pulse sequence shown in Fig.~\ref{fig:pi}. The scheme is organized as two nested loops (Fig.~\ref{fig:loop_protocol}): a fast inner entanglement loop and a slower outer correction loop.

\subsection{Inner entanglement generation loop}
\label{app:innerloop}
\begin{figure}
    \centering
    \includegraphics[width=\figscale\linewidth]{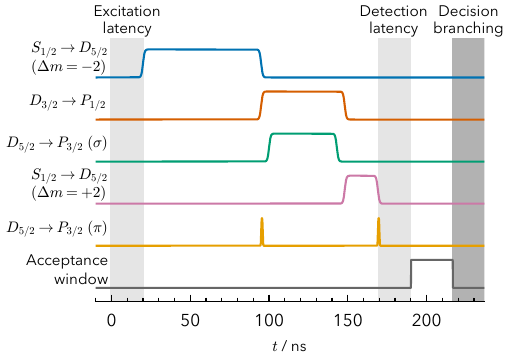}
    \caption{Pulse timing of the 237 ns entanglement generation loop with 20 ns optical latency, 75 ns narrow-line \tpi-time, 50 ns repumping time, \(4\tau\) photon acceptance window, and 20 ns decision time. The latency assumes up to 4 m distance in fiber between the central station and either QPU. The entire loop can be repeated at \SI{4.2}{\mega\hertz}.}
    \label{fig:fastloop}
\end{figure}

Each entanglement attempt consists of a fixed sequence of state preparation, excitation of \zerog~with probability \(p_e\), and decision branching. Figure~\ref{fig:fastloop} shows a graphical representation of the loop timing.

\subsubsection{State preparation}
\begin{figure*}[ht]
    \centering
    \includegraphics[width=0.75\linewidth]{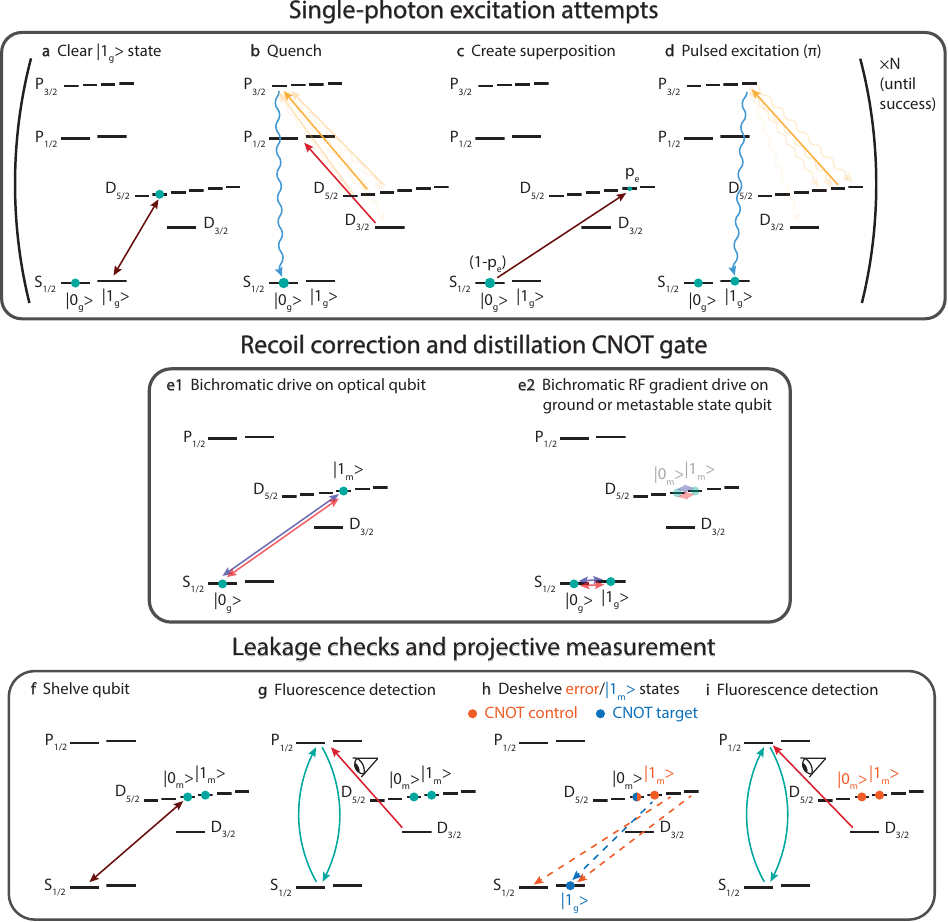}
    \caption{This entanglement and correction pulse sequence is applied simultaneously to ions in two different modules to herald Bell pairs, check for leakage events, and distill higher fidelity Bell pairs.
    Steps \textbf{a,\,b} (optical pumping) and \textbf{c,\,d} (weak excitation and single-photon detection) are repeated until a successful detection event heralds an approximate Bell state.
    The photon recoil correction and two-qubit gate for distillation both use a spin-dependent force that could be supplied by either a laser (\textbf{e1}) or an RF gradient (\textbf{e2}).
    Two fluorescence detection steps project out any errors due to leakage out of the qubit subspace. The second step also reads out the parity of the CNOT target ion for projective distillation (\textbf{f-i}).}
    \label{fig:pi}
\end{figure*}

At the start – when an ion first enters the entanglement loop, or after a previous unsuccessful iteration – the electron population is pumped into \zerog~as quickly as possible to maximize the overall entanglement generation rate. Population remaining in the D\(_{3/2}\) manifold and \oneg~is turned into small rate loss by the subsequent outer-correction-loop checks. However, residual population in the D\(_{5/2}\) manifold could be excited by the \tpi-polarized $\mathrm{D}_{5/2} \rightarrow \mathrm{P}_{3/2}$ pulse and cause a dephasing error which the distillation would not project out if it ends up in \zerog~(App.~\ref{app:distillation}). A convenient implementation using integrated optics, where pure circular polarization presents additional challenges~\cite{Corsetti2026_PolarizationGradientCooling,Clements2026_SubDopplerPolarizationGradient}, is to use a \tpi~pulse on the $\oneg = \ket{\mathrm{S}_{1/2}, m=\frac{1}{2}} \rightarrow \ket{\mathrm{D}_{5/2}, m = -\frac{3}{2}}$ transition, followed by a ``quench'' on the $\mathrm{D}_{5/2}\rightarrow \mathrm{P}_{3/2}$ transition (Fig.~\ref{fig:pi}). Note that as the protocol is operated in the weak excitation regime, $p_e < 1/3$, the steady state benefits from effectively applying the state preparation sequence $p_e^{-1}$ times per excitation out of \zerog.

\subsubsection{Excitation}
Anticipating the outer-loop distillation stage, we excite in a two-step process via the metastable $D_{5/2}$ level, such that angular momentum prevents an excited electron from quickly returning to the \zerog state. First, a narrow-line laser drives a transition between the $\zerog = \ket{\mathrm{S}_{1/2}, m=-\frac{1}{2}}$ and $\ket{\mathrm{D}_{5/2}, m=+\frac{3}{2}}$ states with pulse area \(\theta\) such that \(p_e=\sin^2(\theta/2)\). Subsequently, a short (\(t_\mathrm{p}\ll\tau\)) pulse with \tpi~polarization and \tpi~pulse area excites the $\ket{\mathrm{D}_{5/2}, m=+\frac{3}{2}}$ state to the $\ket{\mathrm{P}_{3/2}, m=+\frac{3}{2}}$ state with lifetime \(\tau\), which decays with high probability to the \oneg{} state and cannot decay back to \zerog. To avoid ending up in \zerog~after excitation, the polarization of the $\mathrm{D}_{5/2} \rightarrow \mathrm{P}_{3/2}$ transition must be \tpi~with high purity. For this excitation scheme, the collection optic should maximize \tpsig-polarized collection. 

\subsubsection{Decision branching and latency}
The $\mathrm{P}_{3/2} \rightarrow \mathrm{S}_{1/2}$ emission from the two nodes is routed to a beamsplitter, where a click in one of the detectors in the output ports heralds the successful preparation of a raw Bell pair. Which detector clicked determines the sign of the entangled state. This sign difference could be tracked as a Pauli frame or removed with a local single-qubit operation. On success, a fast processor must preempt the execution of the next entanglement attempt because further laser pulses would scramble the state. The detectors and control logic are assumed to be located halfway between the two modules to minimize the longest round-trip communication time between control logic and a module, hence optimizing the attempt rate. Here, the relevant latency is the duration between triggering the next attempt and the arrival of the first optical pulse at the ions or conversely the time between photon emission and registering the detector click. Signal propagation in optical fiber sets a practical limit of approximately 5 ns/m for a given node distance, resulting in a 10 ns/m round-trip latency lower bound on the inverse attempt rate unless the excitation loop itself is pipelined.

\subsection{Outer correction loop}
\label{app:outer_loop}
In the correction loop, we apply error mitigation techniques outside the latency-critical entanglement attempt path, merge the two raw Bell pairs to suppress errors through distillation, and deliver successfully entangled qubits to the rest of the processor while re-using the rest of the ions for subsequent rounds. To optimize throughput, it can be implemented by transporting ions between five physically separated pipeline zones (Fig.~\ref{fig:pipeline}). To start, ions are cooled to remove built-up motional excitation in a cooling buffer zone. In the next zone, photon-mediated entanglement is generated (executing the inner entanglement loop), and following a successful herald, residual qubit-motion entanglement is removed. Ions entangled this way are buffered and forwarded in pairs to a zone where a CNOT gate is applied to implement the single-stage entanglement distillation, followed by a fluorescence readout zone for parity and leakage detection. The photon timestamps and the number of attempts before success provide soft information on the ion temperature, phase error, dark count probability, recoil correction, and photon distinguishability that a higher-level decoder could use to mitigate infidelity~\cite{Li2024_NeutralAtomCavityInterconnects}. Assuming that per channel four ions are being transported and cooling-buffered, one ion is in the entanglement zone, four are held in the entanglement buffer, and two ions each are in the distillation and readout zones, then each channel contains 13 ions at steady state.

\subsubsection{Coherent disentanglement of spin and motion}
\label{app:motion}

The phase sensitivity of the Bell states generated by the single-photon scheme requires attention in our architecture for two distinct reasons. Slow drifts in the cross-module link phase are addressed by a single stage of distillation, as described in the next section. However, the motion of the ion emitters in the trap potential during the emission process also scrambles the state phase; this can be viewed as unwanted spin-motion entanglement, and is not reduced by the single distillation stage.

The loss in fidelity due to this effect could be reduced to some extent by orienting the excitation laser along the collection axis (at least for single-stage excitation and low trap frequencies)~\cite{Cabrillo1999_CreationEntangledStates,Liu2026_LongLivedRemoteIonRepeaters}. We discard this option here, as scalable fabrication using surface-electrode traps and integrated photonics strongly favors excitation and collection optics that lie in the same plane. However, with knowledge of the excitation and emission geometry, photon timing, and trap frequencies, the spin-motion entanglement resulting from the excitation and decay process can be calculated exactly and largely undone by a spin-dependent displacement operation~\cite{Apolin2025_RecoilInducedErrors}.

A simple spin-dependent displacement is the exact disentangling correction for the case of instantaneous impulsive excitation and collection from an infinitesimally small solid angle, where it completely removes motional loss of fidelity for arbitrary initial states of motion. For the example \sreight{} parameters considered here, and collection from a Gaussian mode with a hard stop at numerical aperture 0.4 (collection cone half-angle \SI{24}{\degree}), the residual error remains below $10^{-4}$ for thermal states with mean occupation numbers $\bar{n}_{th} \lesssim 20$ and a 2 MHz mode frequency, as shown in Fig.~\ref{fig:fidelity_panels_motion}.

This relative insensitivity to motional effects enables high-rate operation, as no cooling is necessary within an inner entanglement attempt loop. We can estimate an upper bound for the motional excitation in each mode due to the photons scattered during state preparation and excitation as $8/3 |\eta|^2$, where $\eta$ is the Lamb–Dicke parameter for the $\SI{408}{\nano\meter}$ photons scattered. Here, we have assumed that the attempt rate is not commensurate with the motional mode frequency, limiting coherent build-up of excitations. Given an average of $p_e / p_\mathrm{herald}$ excitation events per successful Bell-pair generation and a 2 MHz secular frequency, this loose bound yields a heating estimate of $\approx 1$ quantum due to the entanglement generation process.
Thus, heating due to the entanglement generation process can be neglected for the system-level analysis; the details of the realization of the spin-dependent forces used for the disentangling correction and the subsequent CNOT gate may drive these requirements.

As is standard in trapped-ion two-qubit gates, the spin-dependent displacement can be implemented by a bichromatic drive on the motional sidebands of a resolved sideband transition. Because the recoil displacement is relatively small (magnitude determined by the Lamb–Dicke parameter \(\eta\ll1\)), recoil correction requires less time and precision for equivalent fidelity than a two-qubit gate. The motional frequencies in each trap need to be calibrated locally to a precision given by the latency between scattering event and application of the spin-dependent correction, but are not required to match between nodes.

The two-step excitation process in our scheme requires a slight extension of the treatment in Ref.~\cite{Apolin2025_RecoilInducedErrors}. The second, fast step can still be modeled as an impulsive plane-wave absorption event (though of a different Lamb–Dicke parameter to the emitted photon); the preceding transfer from $\zerog$ on the narrow-line transition is only approximately impulsive, but numerical calculations show that a spin-dependent post-herald displacement nevertheless disentangles spin and motion to $<\num{e-7}$ residual infidelity.

\subsubsection{Distillation}
\label{app:distillation}

\begin{figure}
    \includegraphics[scale=0.8]{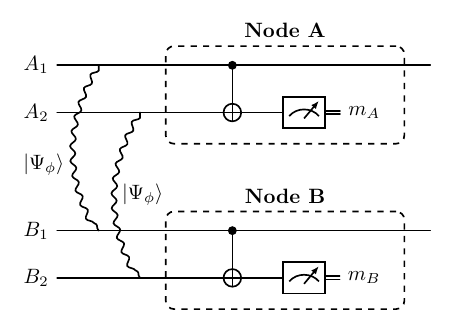}
    \caption{Distillation circuit. For two input Bell pairs $A_1 \otimes B_1$ and $A_2 \otimes B_2$ of the form $\ket{\Psi_{\phi}}=(\ket{10}+e^{i\phi}\ket{01})/\sqrt{2}$, the classical outcomes $(m_A, m_B) = (1, 1)$ of the measurement on the second signal the generation of the fixed output state $\ket{\Psi^+}$ irrespective of the arbitrary, constant phase $\phi$. The circuit also rejects an imbalance between the $\ket{01}$ and $\ket{10}$ components, and an admixture of a $\ket{11}$ component (see text).}
    \label{fig:distillation-circuit}
\end{figure}

In each of the modules, A and B, we combine two subsequent raw Bell pairs (A1, B1 and A2, B2) locally to implement entanglement distillation. We perform a CNOT gate between the qubits, and subsequently read out the target qubit in the Z basis, realizing a parity measurement as proposed by Bennett et al.~\cite{Bennett1996_PurificationNoisyEntanglement} (Fig.~\ref{fig:distillation-circuit}). As pointed out by Campbell and Benjamin~\cite{Campbell2008_ExtremePhotonLoss} (there formulated in a measurement-based setting, but the protocol is equivalent), this is an excellent fit to the errors inherent in single-photon entanglement, rejecting double excitation $\ket{11}$ errors, as well as unknown entangled state phase $\phi$ and collection/beamsplitter asymmetry $\chi$ as long as they are stable across two subsequent Bell pairs. This protocol was experimentally demonstrated in single-photon-based entanglement of nitrogen vacancy center qubits by Kalb et al.~\cite{Kalb2017_EntanglementDistillationNV}.

Concretely, for a raw input Bell pair described by
\begin{equation}
    \begin{aligned}
        \rho_\mathrm{raw} &= (1-p_e)\proj{\Psi_{\chi,\phi}} + p_e \proj{11},\\ 
        \ket{\Psi_{\chi,\phi}}&=\sqrt{\frac{1+\chi}{2}}\ket{10}+e^{i\phi}\sqrt{\frac{1-\chi}{2}}\ket{01},
    \end{aligned}
    \label{eq:distillation_input_state}
\end{equation}
the result (1,1) for measuring the target pair in the Z basis projects the control pair into a pure Bell state \(\ket{\Psi^+}=(\ket{10}+\ket{01})/\sqrt{2}\) independent of the input phase $\phi$, the asymmetry \(\chi\), and the admixture $p_e$ of $\ket{11}$. The other outcomes do not contain entanglement without knowledge of $\phi$, leading to the correction loop being restarted without a Bell pair being delivered to the QPU. The desired outcome occurs with probability $P_{11}=\frac{1}{2}\,(1-\chi^2)\,(1-p_e)^2$ (Eq.~\ref{eq:dist_success}),
leading to optimal performance for symmetric efficiencies $\chi = 0$. The asymmetry infidelity for a single Bell pair \(i\) is \(\frac{(1-\sqrt{1-\chi_i^2})(1-p_e)}{2}\) and the residual infidelity after distillation is \(\approx\frac{(\chi_2-\chi_1)^2}{4}\), set by the asymmetry difference between the two Bell pairs.
If the rate of distillation attempts is dominated by entanglement generation (linear in the excitation efficiency $p_e$), the optimal success rate is achieved at $p_e = 1/3$; if the rate is dominated by another step, the optimal excitation efficiency will be lower to trade off entanglement-loop success probability for increased distillation $P_{11}$. For the $p_e = 0.198$ working point discussed here, distillation succeeds in a fraction $P_{11} = 0.322$ of correction-loop cycles.

\begin{figure}[!tb]
    \centering
    \includegraphics[width=\figscale\linewidth]{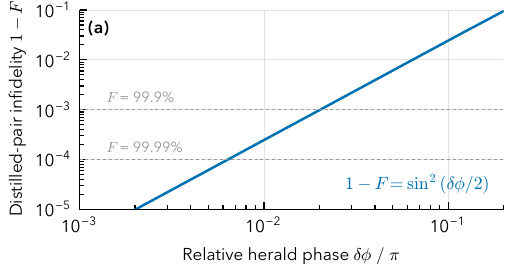}\\[5pt]
    \includegraphics[width=\figscale\linewidth]{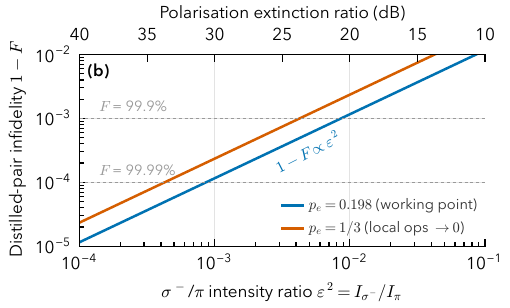}
    \caption{\textbf{(a)} Distilled-pair infidelity to \(\ket{\Psi^+}\) vs.~deterministic relative herald phase \(\delta\phi=\phi_2-\phi_1\) in the balanced case (\(\chi=0\)). \textbf{(b)} Distilled-pair infidelity vs.~the \(\sigma^-/\pi\) drive-intensity ratio \(\varepsilon^2 \equiv I_{\sigma^-}/I_\pi\), for the working-point \(p_e = 0.198\) and the local-operations-free optimum \(p_e = 1/3\); upper axis is the polarization extinction ratio \(\mathrm{PER}=-10\log_{10}\varepsilon^2\).}
    \label{fig:fidelity_panels_phase_and_pol}
\end{figure}

The parity measurement completely rejects all even-parity populations ($\ket{00}$ and $\ket{11}$). Coupled with the herald condition of having detected a photon, such that at least one ion must have emitted a photon, this leads to erroneous population in $\ket{1}$ from imperfect state preparation or additional decay pathways (that do not lead to a signal photon being emitted) being converted into rate loss, which is favorable in the context of a heralded scheme. However, branching back to $\ket{0}$ after excitation is not rejected by the distillation step: a fraction $\epsilon_\mathrm{ret}$ decaying to $\ket{0}$ after excitation in the raw entanglement case leads to a nominal reduction in double excitation infidelity from $p_e$ to $(1- \frac{\epsilon_\mathrm{ret}}{2}) p_e$ (a classical admixture of $\ket{01}$ and $\ket{10}$ contributes $F=1/2$ as opposed to $F=0$ for $\ket{11}$), but after distillation leads to an error of $(1 - (1 + \frac{p_e \epsilon_\mathrm{ret}}{1 - p_e})^{-2})/ 2 = \frac{p_e}{1-p_e} \epsilon_\mathrm{ret} + O(\epsilon_\mathrm{ret}^2)$. As we do not assume the second $\mathrm{D}_{5/2} \rightarrow \mathrm{P}_{3/2}$ step to be spectrally resolved, this yields a requirement on its polarization being purely \tpi{}. For an intensity fraction $\varepsilon^2\equiv I_{\sigma^-}/I_\pi$, the Clebsch–Gordan factors and branching ratio to the ground state set $\epsilon_\mathrm{ret}=\frac{\gamma}{2}\varepsilon^2$ (Fig.~\ref{fig:fidelity_panels_phase_and_pol}b).

While the chosen distillation stage rejects Z rotations common to both Bell pairs, individual Z errors are not rejected.
As two Bell pairs are consumed, this leads to twice the sensitivity to dephasing errors.
Concretely, an additional factor $\mathcal{C}$ ($0 < |\mathcal{C}| < 1$) in the off-diagonal $\ketbra{01}{10}$ coherence element of the raw density matrix (\ref{eq:distillation_input_state}) leads to a factor $|\mathcal{C}|^2$ in the coherence of the post-distillation state.
Thus, the fidelity decreases from $(1-p) (1 + |\mathcal{C}|) / 2$ for a single raw Bell pair to a distilled-state fidelity of $(1 + |\mathcal{C}|^2)/2$ (for small errors, a doubling).
One such dephasing mechanism stems again from the polarization of the $\mathrm{D}_{5/2} \rightarrow \mathrm{P}_{3/2}$ drive: $\ket{P_{3/2},+1/2}$ decays to \oneg{} with probability $2/3$.
If the \tpi-polarized photons emitted by this decay are not entirely filtered out and are sometimes mistaken for \tsig-polarized signal photons, this leads to an admixture of a completely dephased component, reducing the contrast to $\mathcal{C} = (1 - \epsilon_\mathrm{\pi})^2$, where $\epsilon_\mathrm{\pi}$ is the overall probability for such an erroneous photon to be generated and detected per node.
Assuming a drive laser intensity fraction $\varepsilon^2$ (see last paragraph) and that \tpi-polarized photons are filtered out with suppression $\eta_\pi$ gives $\epsilon_\mathrm{\pi}=\frac{1}{2}\eta_\pi\,\varepsilon^2$, yielding a post-distillation error of $\eta_\pi \varepsilon^2 + O((\eta_\pi \varepsilon^2)^2)$.
We neglect this in Fig.~\ref{fig:fidelity_panels_phase_and_pol}(b), as with reasonable polarization selectivity this error will be much smaller than the error due to decays to \zerog.
Other dephasing mechanisms include residual spin–motion entanglement (see App.~\ref{app:motion}), and errors not specific to the optical interface (e.g. qubit frequency noise).

False heralds, where a heralding event occurs independent of the signal photons incident on the beamsplitter network, can occur due to detector dark counts or crosstalk of classical excitation light or photons from other modules. In the limit of low system efficiency and small other errors, it is described by an admixture of a separable state $\rho_\mathrm{false} \otimes \rho_\mathrm{false}$, where $\rho_\mathrm{false} \approx (1-p_e)\proj{0} + p_e \proj{1}$ is the local state after excitation (entirely tracing out the signal photon, if any). As the joint state contains odd-parity terms, the raw infidelity for a certain false herald probability $p_\mathrm{false}$ of $(1-p_e)^2 p_\mathrm{false}$ (on top of the double-excitation error) is imperfectly suppressed by the distillation process to $6 p_e p_\mathrm{false} + O(p_\mathrm{false}^2)$. False heralds merit attention because the otherwise high fidelities inherent to the scheme mean that even 1-s\(^{-1}\) level false heralds may become limiting (especially in passive networks, see App.~\ref{app:networks}). False-herald errors can be traded off against rate using the acceptance window length.

For two Bell pairs (\ref{eq:distillation_input_state}) with generally different phases $\ket{\Psi_{\chi,\alpha}}$ and $\ket{\Psi_{\chi,\beta}}$, the state after successful distillation is $\ket{\Psi_{0, \alpha - \beta}}$. Any common offset $\alpha = \beta = \phi$ is corrected for arbitrary $\phi$, but a difference $\delta \phi$ in phases leads to a loss in fidelity to the nominal output state of $1 - |\braket{\Psi^+|\Psi_{0, \delta\phi}}|^2 = \sin^2 (\delta\phi/2)$ (see App.~\ref{app:phase-stability}).

\subsubsection{Leakage detection}
\label{app:leakage}

There are several leakage states that can be populated after a successful herald due to nonzero branching ratios of the P\(_{3/2}\) level to D\(_{5/2}\) and D\(_{3/2}\) and possible population left in these manifolds after state preparation. By rotating the qubit into the \(m=\pm1/2\) states of the D\(_{5/2}\) manifold, a first round of fluorescence detection checks for leakage errors that were in these states or the D\(_{3/2}\) manifold. If the first fluorescence result is dark, a second round applies different operations to the control and target ions of the distillation CNOT and requires individually resolving the two ions. For the control ion, the second round deshelves the \(m=3/2, 5/2\) states of the D\(_{5/2}\) manifold that could have been populated in the excitation. For the target ion, the qubit state that heralds a successful parity check is deshelved. A second dark fluorescence result on the control ions and bright results on the target ions projects out the remaining leakage errors and signals a successfully distilled Bell pair. If the control ion is measured in a separate potential well, its motion is not heated by the bright measurement of the target ion in successful distillation runs. Separating the leakage checks and parity measurement into three detections makes it possible for a successful distillation to require all-dark detection for both ions and thus minimize motional excitation and recooling.

The two-step excitation allows leakage detection and distillation to catch most imperfect-extinction errors and turn them into rate hits. All possible remaining errors require at least two improbable events; an example error pathway is: excitation to \(P\) (\(p\approx0.2\)), then decay back into the metastable manifold with \(\Delta m \ne0\) after excitation (\(p\approx0.004\) for \sreight) and residual \(D\rightarrow P\) light re-exciting (extinction ratio \(\eta_\mathrm{ex}\)) and the electron decaying back to the initial state (\(p\approx0.16\)), totaling \(p\approx0.0001\eta_\mathrm{ex}\).

\section{Link phase sensitivity}
\label{app:phase-stability}

The phase sensitivity of the entangled states generated by the single-photon scheme can be understood by analogy to a Mach--Zehnder interferometer whose two arms consist of the combined excitation and photon emission pathways at the two nodes.
In our two-step excitation scheme, this includes the optical phase of both laser pulses at the ion, as well as the phase accumulated along the optical path from the collection optic to the beamsplitter at which the photons emitted from the two nodes are interfered.
The distillation scheme removes the requirement for the two paths to be interferometrically stabilized globally and across longer times (App.~\ref{app:distillation}), but the overall phase still needs to be stable at the shorter time scale set by the average inter-herald interval.
Assuming subsequent Bell pairs used for distillation are generated using the same routing configuration, the dominant source of phase fluctuations are likely to be optical fibers connecting the modules.

\newcommand{\avgheraldinterval}{T}
\newcommand{\expect}[1]{\mathbb{E}\left[ #1 \right]}
\newcommand{\dd}[1]{{\mathrm{d}#1}}
\newcommand{\phaseerr}{\epsilon_\textrm{link-phase}}
Consider two successive heralds occurring at times $t_0$ and $t_1$.
Neglecting the discrete nature of the attempt cycles, the inter-herald interval $\delta t = t_1 - t_0$ is well-approximated by an exponential distribution with a mean $\avgheraldinterval = t_\mathrm{cycle} / p_\mathrm{herald}$.
The average infidelity $\phaseerr$ is then (Table~\ref{tab:fidelity}, App.~\ref{app:distillation})
\begin{equation}
    \begin{aligned}
       \phaseerr &= \expect{\sin^2\frac{\delta\phi}{2}}\\
        &= \frac{1}{2}\left(1 - \frac{1}{\avgheraldinterval} \int_{0}^{\infty} e^{-\frac{\delta t}{\avgheraldinterval}} \expect{\cos \delta\phi|\delta t} \dd{\delta t}\right),
    \end{aligned}
\end{equation}
where $\expect{\cos \delta\phi|\delta t} = \operatorname{re}\expect{e^{i \delta\phi}|\delta t}$ is obtained from the characteristic function of the phase increment $\delta \phi$ at a given $\delta t$.
Assuming $\delta\phi$ follows a zero-mean Gaussian distribution~\cite{Minar2008_PhaseNoiseLongFiber}, we have $\expect{\cos \delta\phi|\delta t} = e^{-\sigma^2_{\!\delta\phi}(\delta t)/2}$, where the phase-increment variance $\sigma^2_{\!\delta\phi}$ may be expressed in terms of the phase noise spectrum $S_\phi(f)$ as
\begin{equation}
    \sigma^2_{\!\delta\phi}(\delta t) = 4 \int_{0}^{\infty} S_\phi(f)\, \sin^2(\pi f \delta t)\, \dd{\hspace{-1pt}f}.
\end{equation}
Here, $S_\phi(f)$ means a power spectral density in the usual engineering sense; the phase evolution may not strictly be a stationary process.
A white-frequency-noise (Wiener) process with $S_\phi(f) = b_{-2} / f^2$ yields $\phaseerr = (1 + (b_{-2\,} \pi^2\, \avgheraldinterval)^{-1})^{-1} / 2$.

In the small-error regime, i.e.~neglecting higher moments (whether Gaussian or not), $\phaseerr \approx \mathbb{E}[\sigma_{\delta\phi}^2(\delta t)] / 4$, and hence
\begin{equation}
    \phaseerr \approx \frac{1}{2} \int_{0}^{\infty} S_\phi(f)\, \frac{1}{1 + 1 / (2 \pi f \avgheraldinterval)^2} \dd{\hspace{-1pt}f}.
\end{equation}
Distillation thus suppresses link phase noise akin to a first-order high-pass filter with a corner frequency given by the inverse of the average inter-herald interval, $1 / (2 \pi \avgheraldinterval)$.

The phase stability of single-mode fiber links depends strongly on their mechanical environment; vibrations typically dominate in the frequency band of interest.
For \SI{10}{\metre}-scale patch cords in a laboratory environment, phase noise densities of $S_\phi(f) \sim 10^{-3} \ldots 10^{-1} / f^2\, \mathrm{rad}^2\, \mathrm{Hz}$ have been reported in the literature~\cite{Grosche2009_OpticalFrequencyTransfer,Mukherjee2022_DigitalDopplerCancellation}.
In Table~\ref{tab:params}, we give a representative phase deviation $\sigma_{\delta\phi} = \sqrt{\mathbb{E}[\sigma_{\delta\phi}^2(\delta t)]} = \SI{0.02}{\radian}$ based on the upper end of that range, equivalent to an error of $\phaseerr = \SI{0.007}{\percent}$.
In experiments optimized for passive stability of fiber links on the $\SI{10}{\metre}$ scale, significantly better stability is routinely achieved~\cite{Hacker2023_PhaseLockingSinglePhoton}, giving rise to negligible estimated infidelities $\phaseerr < \num{e-6}$; the floor due to thermal noise is yet lower~\cite{Bartolo2012_ThermalPhaseNoise}.
On the other hand, local disturbances can dominate the stability independent of the link length, such as observed in Ref.~\cite{Liu2026_LongLivedRemoteIonRepeaters}, where a similar level of stability is reported for \SI{10}{\metre} and \SI{10}{\kilo\metre} links (equivalent to $\phaseerr \approx \SI{0.1}{\percent}$).
In situations where the passive stability of the link phase is not sufficient, active stabilization may be employed.

\section{Photonic networks for excitation and interference}
\label{app:networks}

\begin{figure}
    \centering
    \includegraphics[width=\figscale\linewidth]{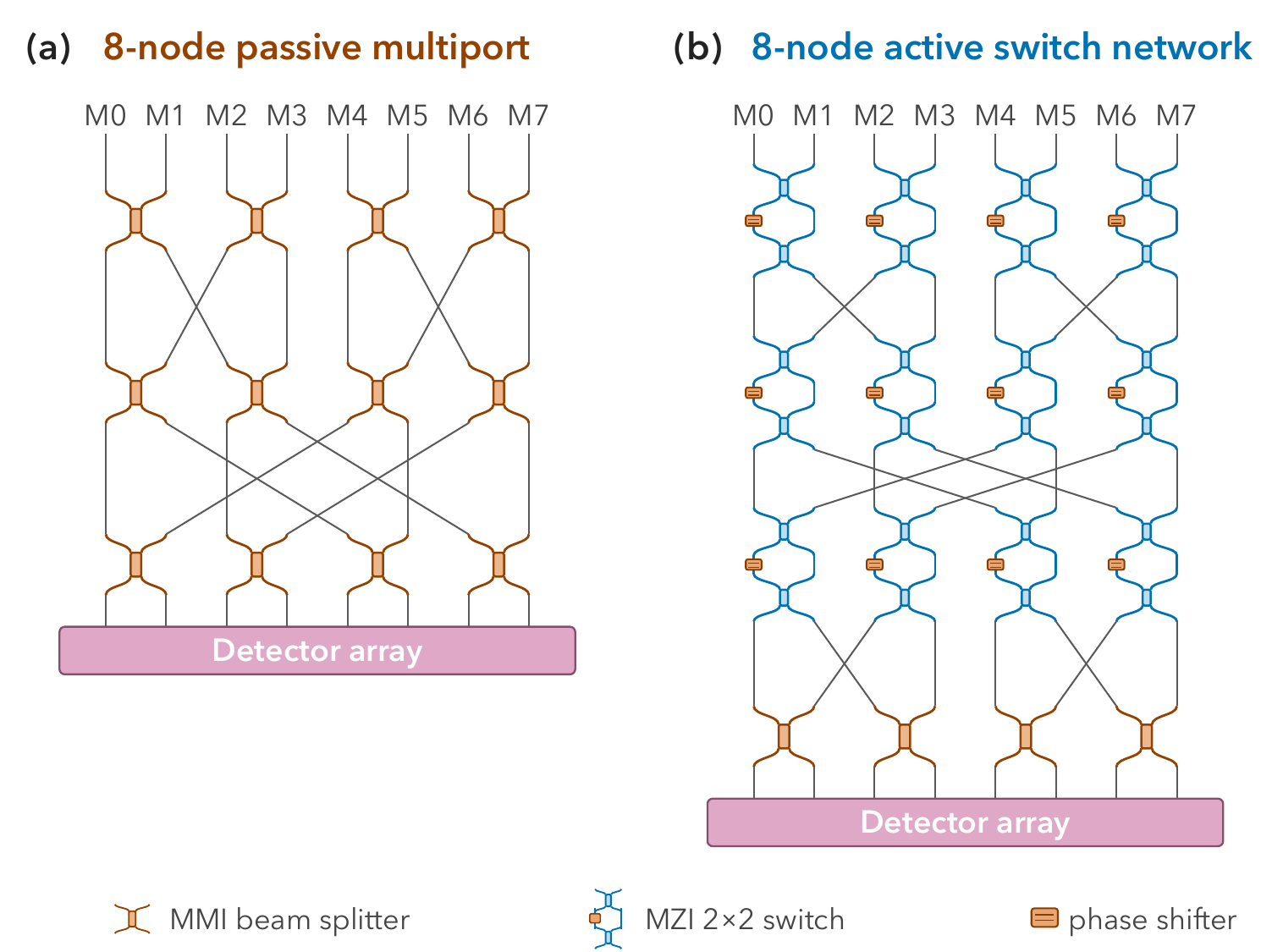}
    \caption{A comparison of the active and passive photonic network for a single channel and eight modules.}
    \label{fig:network_examples}
\end{figure}

On the emission side, the choice of a switch network or a passive symmetric multiport (see Fig.~\ref{fig:network_examples}) carries fidelity consequences as well as parallelism implications. Both designs discussed here assume optical paths can cross, which reduces the element number compared to crossing-free networks~\cite{Koyama2024_OptimalSwitchingNetworks}. While the depth and total element number are comparable, an active network allows fully parallel connection (as shown in the following section), while a passive network only supports entangling two of \(M\) modules at any given time. By contrast, the passive device is simpler, easier to fabricate, requires fewer controls, and may incur less loss. The networks face competing fidelity trade-offs: an active network accrues infidelity due to crosstalk at each switch crossing a live line, whereas the passive case has no network crosstalk because only two inputs are simultaneously active. The infidelity from detector dark counts scales with the number of active detectors per Bell pair and thus favors active networks. For both networks, ignoring heralds from a detector for a short time after it fires mitigates detector afterpulsing errors at low rate cost. 

Both dark-count and crosstalk infidelity are proportional to the probability that a given herald arises from a false-positive detection \(p_\mathrm{false}\). For \(n_d\) detectors with dark-count rate \(R_\mathrm{dc}\),
\begin{equation}
    p_\mathrm{false}^\mathrm{(dc)} = \frac{n_d\,R_\mathrm{dc}\,t_\mathrm{window}}{p_\mathrm{herald}}
\end{equation}
The infidelity is \(\frac{M}{2}\) times larger for the multiport than for the switch network. A second potential source of infidelity is crosstalk arising either from crossed optical signals in waveguides or imperfect switches. Since the symmetric multiport supports only a single active pair at any given time and distillation turns static imbalances into rate hits, crosstalk causes no infidelity in passive networks. In the butterfly switch network, \(p_\mathrm{false}^\mathrm{(X)} = n_X\,\epsilon_X\), the number of opportunities \(n_X\) for other photons to leak into the path times the probability \(\epsilon_X\). Time-interleaving the excitations on different modules could increase the parallelism of passive networks and reduce the crosstalk errors in switch networks while also making more efficient use of laser power. However, an additional switch may be required per input module to gate photons scattered during state preparation.

\begin{table}[t]
    \centering
    \begin{ruledtabular}
    \begin{tabular}{l c c}
        & Passive & Active \\ [1pt]
        \hline\\ [-8pt]
        Switches                  & 0                              & \(\tfrac{M}{2}\log_2 M\) \\
        Beamsplitters            & \(\tfrac{M}{2}\log_2 M\)       & \(\tfrac{M}{2}\) \\
        Detectors                 & \(M\)                          & \(M\) \\
        Depth                     & \(\log_2 M\) BS                & \(\le\log_2 M\) SW\,+\,1 BS \\
        Simultaneous connections  & 1 pair                         & all \(\tfrac{M}{2}\) pairs \\[2pt]
        \hline\\ [-8pt]
        Crosstalk infidelity      & ---                            &  \(\le\log_2 M\) factors \\
        Detectors per pair        & \(M\)                        & 2 \\
    \end{tabular}
    \end{ruledtabular}
    \caption{Comparison of passive and active photonic networks for
    realizing all-to-all connectivity between modules. Counts assume
    \(M=2^D\) with \(D=\lceil\log_2 \mathrm{modules}\rceil\). BS = beamsplitter, SW = switch.}
    \label{tab:photonic_networks}
\end{table}

\subsection*{The butterfly network routes any pairing}

We show that, for any pairing of the $M=2^D$ modules, a depth-$D$
butterfly network can route the two members of every pair to a
beamsplitter of their own.

We label each output port with a $D$-bit address, where the $D-1$ higher
bits give the binary expansion of the beamsplitter index. The low bit
labels which of the two ports of that beamsplitter the module occupies,
and we call it the beamsplitter-port bit. We number the stages so that
stage $0$ acts on the module inputs and stage $D-1$ precedes the fixed
beamsplitter layer. A stage-$s$ switch separates its inputs according to
bit $s$ of the target address. It can be set correctly if and only if
its two inputs require opposite values of that
bit~\cite{Benes1965_ConnectingNetworks}.

The routing condition at each stage thus becomes a binary assignment
problem: assign one bit to each object in a collection, with certain
pairs of objects required to receive opposite values. In the stages
considered here, these requirements always come from two pairings of
the same collection. Drawing both pairings as edges gives every object
one edge of each kind, so the graph separates into cycles that
alternate between the two edge types and are therefore even. Assigning
bits $0,1,0,1,\ldots$ around each cycle then satisfies every
requirement of both pairings at once. This alternating construction is
the cycle-decomposition step of the classical looping algorithm for
rearrangeable networks~\cite{Benes1965_ConnectingNetworks}. All that
remains is to identify the two pairings at each stage.

Stage $0$ sets the beamsplitter-port bit of every module. This bit must
differ between the two members of a matched pair, as they must occupy
the two ports of one beamsplitter. Take the desired connections between
modules as one pairing, and the hard-wired input pairing of modules entering
the same stage-$0$ switch as the other. By the observation above, the
port bits can be chosen to differ both between the members of every
matched pair and between the two inputs of every stage-$0$ switch.
Every stage-$0$ switch can then be set correctly.

Each later stage fixes one bit of the shared beamsplitter index. Assume
stages $0,\ldots,s-1$ have been set correctly, meaning each module occupies a
position whose address agrees with its target in all bits below $s$. A
stage-$s$ switch joins positions differing only in bit $s$, so its two
inputs agree on all lower bits, including the beamsplitter-port bit.
Modules with port bit $0$ therefore meet only other modules with port
bit $0$ (likewise for port bit $1$).

The same pairing argument now applies with the matched pairs themselves
as the objects. The port-$0$ modules, one per matched pair, are paired
up by which stage-$s$ switch they enter. No pair meets itself in this
pairing, as each pair contributes only one port-$0$ module. The
port-$1$ modules define a second pairing in the same way. Alternating
the stage-$s$ beamsplitter-index bit around the resulting cycles gives
one shared bit to each matched pair and opposite bit values to the two
modules entering every stage-$s$ switch. Every stage-$s$ switch can then
be set correctly. Continuing through $s=1,\ldots,D-1$ fixes the full
beamsplitter index of every pair, and every module reaches its target
address. Because the network of two-port switches implements a permutation, every
output port receives exactly one module, so no two pairs share a
beamsplitter.

\section{Narrow-line power requirement}
\label{app:power}
The rate model in Sec.~\ref{sec:rates} treats the narrow-line \(\pi\) time \(t_\pi\) as an input parameter. Here we translate that parameter into an optical power per entanglement channel. The dipole-allowed \(D\to P\) step requires orders of magnitude less power at the same waist and is neglected below.

\begin{figure}
    \centering
    \includegraphics[width=\figscale\linewidth]{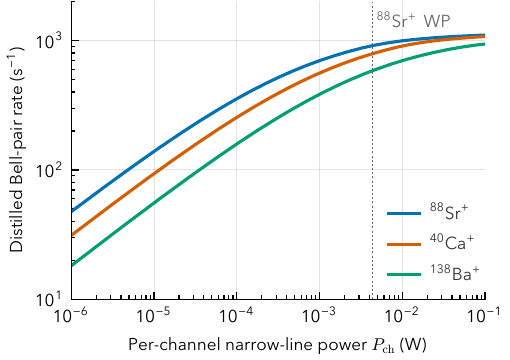}
    \caption{Single-excitation distilled Bell-pair rate vs.\ per-channel narrow-line power \(P_\mathrm{ch}\) for \sreight, \cafour, and \baeight. Assuming a rectangular pulse and \(1/e^2\) diameter \(d=\SI{2}{\micro\meter}\) Gaussian beams, with \(p_e\) re-optimized at each power. The narrow-line \(\pi\)-time follows from Eq.~\ref{eq:tpi_power} using \(\tau_{D_{5/2}}=0.39,\,1.17,\,31\,\mathrm{s}\) and wavelengths \(674,\,729,\,1762\,\mathrm{nm}\) (Sr\(^+\)/Ca\(^+\)/Ba\(^+\)); the remaining cycle-time and herald-probability inputs are \(\tau=6.6,\,6.6,\,6.3\,\mathrm{ns}\) and \(\gamma=0.94,\,0.935,\,0.74\), with other parameters following Table~\ref{tab:params}. The dashed line marks the \sreight~working point (\(t_\pi=75\,\mathrm{ns}\), \(P_\mathrm{ch}\approx4.4\,\mathrm{mW}\)).}
    \label{fig:rate_vs_power}
\end{figure}

For a resonant quadrupole transition driven by a spatial Gaussian beam with \(1/e^2\) intensity diameter \(d\) and power \(P_\mathrm{ch}\),
\begin{equation}
    t_\pi(P_\mathrm{ch},d) = d\,\sqrt{\frac{\pi^5\,\hbar c\,\tau_{D_{5/2}}}{60\,\lambda^3\,C_\mathrm{eff}^2\,P_\mathrm{ch}}}\,,
    \label{eq:tpi_power}
\end{equation}
where \(\tau_{D_{5/2}}\) is the upper-state radiative lifetime and \(C_\mathrm{eff}^2=1/30\) is the product of the Clebsch--Gordan (\(1/5\)) and geometric (\(1/6\)) factors for our \(|\Delta m|=2\) drive at the optimal beam/field orientation~\cite{Roos2000_ControllingQuantumState}. For \sreight, Eq.~\ref{eq:tpi_power} gives \(t_\pi=75\,\mathrm{ns}\) at \(P_\mathrm{ch}\simeq1.1\), \(4.4\), and \(9.9\,\mathrm{mW}\) for \(d=1\), \(2\), and \(\SI{3}{\micro\meter}\). The rate rises with power while \(t_\pi\) limits the cycle time, then flattens near \(1\,\mathrm{kHz}\) once \(t_\pi\) becomes negligible. Among the species considered in Fig.~\ref{fig:rate_vs_power}, \sreight~reaches a given rate with the least power because of its favorable quadrupole matrix element and transition wavelength.

\section{Resource requirements}
\label{app:requirements}
In this section we explain the assumptions behind Table~\ref{tab:requirements_area} and, more importantly, show that upstream choices set the three quantities the interconnect must deliver. The QEC encoding (transversal surface code vs.\ qLDPC code with surgery) and the syndrome measurement cycle time \(\tau_\mathrm{syndrome}\) govern the Bell-pair rate and the algorithm run time, while the style of the remote operation (direct Bell-pair-mediated gates vs.\ consuming distilled logical Bell pairs) governs the Bell-pair fidelity required. The details of the codes are out of scope; we simply adopt the codes that recent architecture papers use to solve large-scale problems such as factoring a 2048-bit RSA number or quantum chemistry~\cite{Zhou2025_TransversalArchitecture,Cain2026_ShorReconfigurableAtoms}, and, as a representative scale, derive the number of Bell pairs each QPU must supply to sustain one remote two-qubit logical operation between a pair of QPUs every logical cycle~\cite{Tripier2026_WalkingCatArchitecture}.

The encoding fixes how many Bell pairs a remote logical operation costs and how that cost is spread in time. A surface-code patch of distance \(d\) has \(d^2\) data qubits, so a transversal CNOT between two patches on different modules consumes \(d^2\) Bell pairs in a single syndrome-extraction round~\cite{Zhou2025_TransversalArchitecture}. A CNOT between two qLDPC blocks instead proceeds by a joint Pauli measurement with a bridge system~\cite{Swaroop2025_UniversalQLDPCAdapters,He2025_QLDPCExtractors,Cross2025_ImprovedQLDPCSurgery}, which consumes \(O(d)\) Bell pairs per round but must be repeated for \(O(d)\) rounds~\cite{Swaroop2025_UniversalQLDPCAdapters,Yoder2025_TourDeGrossBicycle}. The two therefore sit at opposite ends of a rate--run-time trade-off: for a comparable total Bell-pair count, the transversal scheme requires \(O(d)\) more Bell pairs per round, but has \(O(d)\) shorter run time. The required Bell-pair rate is the per-round pair count divided by \(\tau_\mathrm{syndrome}\), so the aggressive \(\tau_\mathrm{syndrome}=1\)~ms assumed by architecture papers~\cite{Zhou2025_TransversalArchitecture,Webster2026_PinnacleArchitecture,Cain2026_ShorReconfigurableAtoms} and the near-term \(\tau_\mathrm{syndrome}=\SI{55}{\milli\second}\) of the Quantinuum Helios operations layer~\cite{Ransford2025_Helios} already span more than an order of magnitude in Bell-pair rate.

The distance needed, and hence the absolute Bell-pair count, grows only slowly with problem size. Treating idle, local, and remote operations on equal footing through the circuit spacetime volume \(V\) (a conservative choice), and assuming a sub-threshold logical error \(p_L\approx \kappa\,(\frac{p}{p_{th}})^{\frac{d+1}{2}}\) with physical error rate \(p\), threshold \(p_{th}\), and constant \(\kappa\), the minimum distance for which the total logical error stays of order one is
\begin{equation}
    d\approx\frac{2\ln(\kappa\,V)}{\ln(p_{th}/p)}-1.
    \label{eq:min_distance}
\end{equation}
Because the distance grows only logarithmically with circuit volume, \(d\propto\ln(\kappa\,V)\), so does the number of Bell pairs consumed per syndrome cycle, through its dependence on \(d\): it is \(\propto d\propto\ln(\kappa\,V)\) for surgery and \(\propto d^2\propto\ln^2(\kappa\,V)\) for transversal gates. The concrete cases of Table~\ref{tab:requirements_area} use distance-\(27\) surface-code patches (\(d^2=729\) Bell pairs in one round)~\cite{Zhou2025_TransversalArchitecture} or a smaller distance \(d\le20\) lifted-product code for surgery~\cite{Cain2026_ShorReconfigurableAtoms}, consuming approximately \(2d=40\) to \(4d=80\) Bell pairs per round depending on the adapter~\cite{Swaroop2025_UniversalQLDPCAdapters} (but assuming a higher operation fidelity of \(\ge99.93\%\) vs.\ \(\ge99.9\%\)).

The structure of the remote operation sets the required Bell-pair fidelity. Both codes above assume approximately \(\ge99.9\%\)-fidelity operations to run RSA-2048 factoring and comparable algorithms; however, for certain operation structures the link fidelity could be lower than the bulk fidelity. Bell-pair-mediated lattice surgery between two surface-code patches tolerates \(14\times\) higher error on the seam than on local operations within a patch~\cite{Ramette2024_FaultTolerantNoisyLinks} (to our knowledge, the analogous tolerance for a bridge between two qLDPC surgery gadgets is unexplored). As an alternative approach that we do not analyze further, two \([[n,k,d]]\) qLDPC blocks in different QPUs can be entangled into \(k\) logical Bell pairs by consuming \(n\) physical Bell pairs, either through a transversal CNOT when the code is CSS~\cite{Li2024_NeutralAtomCavityInterconnects} or through distillation~\cite{BonillaAtaides2025_ConstantOverheadBellDistillation}. The distilled logical pairs can be built from lower-fidelity physical pairs at the cost of more physical qubits and extra distillation code blocks per connected module.

Reaching the fast-cycle rates summarized in Table~\ref{tab:requirements_area} will require multiplexing many Bell-pair generation channels that are dense enough to supply the Bell factory within the cm\(^2\) scale of a single module.

\newpage
\ifpreprint
  \bibliographystyle{aipnum4-1}
\else
  \bibliographystyle{apsrev4-2}
\fi
\bibliography{references}

\end{document}